\documentclass[twocolumn]{aastex631}

\usepackage{color}

\begin{document}

\title{Evolution of the Size-Mass Relation of Star-forming Galaxies Since $z=5.5$ Revealed by CEERS}

\author{Ethan Ward}
\affiliation{Dept. of Physics \& Astronomy, University of California, 900 University Ave, Riverside, CA 92521 USA}

\author[0000-0002-6219-5558]{Alexander de la Vega}
\affiliation{Dept. of Physics \& Astronomy, University of California, 900 University Ave, Riverside, CA 92521 USA}

\author[0000-0001-5846-4404]{Bahram Mobasher}
\affiliation{Dept. of Physics \& Astronomy, University of California, 900 University Ave, Riverside, CA 92521 USA}

\author[0000-0001-8688-2443]{Elizabeth J. McGrath}
\affiliation{Department of Physics and Astronomy, Colby College, Waterville, ME 04901, USA}

\author[0000-0001-9298-3523]{Kartheik G. Iyer}
\altaffiliation{Hubble Fellow}
\affiliation{Columbia Astrophysics Laboratory, Columbia University, 550 West 120th Street, New York, NY 10027, USA}

\author[0000-0003-2536-1614]{Antonello Calabr{\`o}} 
\affiliation{INAF - Osservatorio Astronomico di Roma, via di Frascati 33, 00078 Monte Porzio Catone, Italy}

\author[0000-0001-6820-0015]{Luca Costantin}
\affiliation{Centro de Astrobiolog\'ia (CAB), CSIC-INTA, Ctra de Ajalvir km 4, Torrej\'on de Ardoz, 28850, Madrid, Spain}

\author[0000-0001-5414-5131]{Mark Dickinson}
\affiliation{NSF's National Optical-Infrared Astronomy Research Laboratory, 950 N. Cherry Ave., Tucson, AZ 85719, USA}

\author[0000-0002-4884-6756]{Benne W. Holwerda} 
\affiliation{Department of Physics and Astronomy, University of Louisville, Louisville KY 40292, USA} 

\author[0000-0002-1416-8483]{Marc Huertas-Company}
\affil{Instituto de Astrof\'isica de Canarias, La Laguna, Tenerife, Spain}
\affil{Universidad de la Laguna, La Laguna, Tenerife, Spain}
\affil{Universit\'e Paris-Cit\'e, LERMA - Observatoire de Paris, PSL, Paris, France}

\author[0000-0002-3301-3321]{Michaela Hirschmann}
\affiliation{Institute of Physics, Laboratory of Galaxy Evolution, Ecole Polytechnique Fédérale de Lausanne (EPFL), Observatoire de Sauverny, 1290 Versoix, Switzerland}

\author[0000-0003-1581-7825]{Ray A. Lucas}
\affiliation{Space Telescope Science Institute, 3700 San Martin Drive, Baltimore, MD 21218, USA}

\author[0000-0002-2499-9205]{Viraj Pandya}
\altaffiliation{Hubble Fellow}
\affiliation{Columbia Astrophysics Laboratory, Columbia University, 550 West 120th Street, New York, NY 10027, USA}

\author[0000-0003-3903-6935]{Stephen M.~Wilkins} %
\affiliation{Astronomy Centre, University of Sussex, Falmer, Brighton BN1 9QH, UK}
\affiliation{Institute of Space Sciences and Astronomy, University of Malta, Msida MSD 2080, Malta}

\author[0000-0003-3466-035X]{{L. Y. Aaron} {Yung}}
\altaffiliation{NASA Postdoctoral Fellow}
\affiliation{Astrophysics Science Division, NASA Goddard Space Flight Center, 8800 Greenbelt Rd, Greenbelt, MD 20771, USA}

\author[0000-0002-7959-8783]{Pablo Arrabal Haro}
\affiliation{NSF's National Optical-Infrared Astronomy Research Laboratory, 950 N. Cherry Ave., Tucson, AZ 85719, USA}

\author[0000-0002-9921-9218]{Micaela B. Bagley}
\affiliation{Department of Astronomy, The University of Texas at Austin, Austin, TX, USA}

\author[0000-0001-8519-1130]{Steven L. Finkelstein}
\affiliation{Department of Astronomy, The University of Texas at Austin, Austin, TX, USA}

\author[0000-0001-9187-3605]{Jeyhan S. Kartaltepe}
\affiliation{Laboratory for Multiwavelength Astrophysics, School of Physics and Astronomy, Rochester Institute of Technology, 84 Lomb Memorial Drive, Rochester, NY 14623, USA}

\author[0000-0002-6610-2048]{Anton M. Koekemoer}
\affiliation{Space Telescope Science Institute, 3700 San Martin Dr., Baltimore, MD 21218, USA}

\author[0000-0001-7503-8482]{Casey Papovich}
\affiliation{Department of Physics and Astronomy, Texas A\&M University, College Station, TX, 77843-4242 USA}
\affiliation{George P.\ and Cynthia Woods Mitchell Institute for Fundamental Physics and Astronomy, Texas A\&M University, College Station, TX, 77843-4242 USA}

\author[0000-0003-3382-5941]{Nor Pirzkal}
\affiliation{ESA/AURA Space Telescope Science Institute}




\begin{abstract}
We combine deep imaging data from the CEERS early release {\it JWST} survey and {\it HST} imaging from CANDELS to examine the size-mass relation of star-forming galaxies and the morphology-quenching relation at stellar masses $\textrm{M}_{\star} \geq 10^{9.5} \ \textrm{M}_{\odot}$ over the redshift range $0.5 < z < 5.5$. In this study with a sample of 2,450 galaxies, we separate star-forming and quiescent galaxies based on their star-formation activity and confirm that star-forming and quiescent galaxies have different morphologies out to $z=5.5$, extending the results of earlier studies out to higher redshifts. We find that star-forming and quiescent galaxies have typical S\'{e}rsic indices of $n\sim1.3$ and $n\sim4.3$, respectively.  Focusing on star-forming galaxies, we find that the slope of the size-mass relation is nearly constant with redshift, as was found previously, but shows a modest increase at $z \sim 4.2$. The intercept in the size-mass relation declines out to $z=5.5$ at rates that are similar to what earlier studies found. The intrinsic scatter in the size-mass relation is relatively constant out to $z=5.5$. 
\end{abstract}
\keywords{Galaxy evolution(594) --- High-redshift galaxies(734) --- Galaxy photometry(611) --- Galaxy structure(622)}


\section{Introduction} \label{sec:intro}
The size evolution of the stellar components of galaxies can reveal their assembly histories and how they relate to their host dark matter halos \citep{Fall80, MMW98, Kravtsov13}. For example, mergers of gas-poor galaxies often lead to larger systems. While major mergers yield proportionately larger remnants, minor mergers deposit more materials in the outskirt of galaxies leading to faster growth of galaxies \citep{Bezanson09, Naab09}. The result of this is that by including gas, size evolution becomes more complex. Slow, gas-rich mergers lead to larger disks \citep{Robertson06}, with stellar feedback causing changes in the size independent of mergers, as a result of gaseous outflows modifying galactic potential wells \citep{ElBadry16}. Sizes of galaxies also shed clues on  properties of their dark matter halos. Galaxy sizes may be proportional to virial radii as a consequence of angular momentum conservation during the collapse of a galaxy \citep{MMW98, Dutton07, Kravtsov13}. Furthermore, study of the scatter in galaxy sizes can be used to constrain their halo spins \citep{Bullock01, Maccio08}. 

Observations show that galaxy sizes depend heavily on stellar mass, star-formation activity, and redshift. Holding other galaxy parameters constant, galaxy sizes are larger at lower redshift; star-forming galaxies are larger than quiescent galaxies; and size correlates with stellar mass \citep{Kormendy96, Shen03, Ferguson04, Trujillo06, Buitrago08, Conselice11, Bruce12, Carollo13, Ono13, vdw14, Lange15, Faisst17, Dimauro19, Mowla19, Nedkova21}. At intermediate stellar masses, star-forming galaxies obey a shallow relation between stellar mass and size, $\textrm{R}_{\textrm{eff}} \propto \textrm{M}_{\star}^{0.2}$, where $\textrm{R}_{\textrm{eff}}$ is the half-light radius. For quiescent galaxies, this is much steeper, $\textrm{R}_{\textrm{eff}} \propto \textrm{M}_{\star}^{0.8}$. There is noticeable scatter in the size-mass relations for both types of galaxies \citep{vdw14, Mowla19, Nedkova21}. 

%Broadly speaking, sizes are larger for galaxies that lie at lower redshifts, galaxies that are star-forming, and galaxies that are massive 

Observations also find that star-forming and quiescent galaxies have inherently different morphologies. Star-forming galaxies tend to have light distributions consistent with exponential disks with S\'{e}rsic indices $n\sim1$, while quiescent galaxies tend to have compact light distributions with S\'{e}rsic indices $n\sim4$ \citep{Kauffmann03, Baldry06, Wuyts11}. 

Most of the above studies reached a maximum redshift of $z=3$ due to limitations in the {\it Hubble Space Telescope's} ({\it HST}) sensitivities and angular resolution. Namely, {\it HST} is limited in wavelength such that the rest-frame $V$-band is observable only up to $z=2$. 
Other studies have taken an alternative route by conducting measurements in the rest-frame ultraviolet \citep{Oesch10, Mosleh12, Ono13, Shibuya15, Holwerda15}. However, these studies have conflicted with those made at rest-frame optical wavelengths \citet{vdw14} because galaxies exhibit different morphologies as a function of wavelength \citep[e.g.,][]{Vulcani14}. It is thus crucial to measure morphological parameters at the same rest-frame wavelength across the redshift range of interest. 

Furthermore, in any study involving sizes of galaxies, it is crucial to measure rest-frame sizes in a physically motivated fashion. Studies that examine the size-mass relation in the rest-frame $V$-band typically apply corrections to infer sizes that are statistical in nature, and likely not appropriate on a galaxy-by-galaxy basis \cite[e.g.,][]{vdw14, Mowla19}. In this paper, we apply techniques developed by the MegaMorph project \citep{Barden12, Haussler13, Vika13, Haussler22} to infer rest-frame optical morphologies for individual galaxies using multi-wavelength data. 

From {\it HST} imaging data, there is a hard limit on the highest redshift observable in the rest-frame $V$-band, $z=2$. With the {\it James Webb Space Telescope} ({\it JWST}), however, we can observe the rest-frame $V$-band out to $z\sim7$ using the NIRCam instrument, due to the significantly longer wavelengths probed by {\it JWST} at angular resolution comparable to or better than that of {\it HST}. 

The Cosmic Evolution Early Release Survey \citep[CEERS,][]{Finkesltein22, Bagley23, Finkelstein23} provides a dataset that a number of studies have recently used to examine the morphologies of galaxies at high redshift. \citet{Ferreira23} and \citet{Kartaltepe23} find that disk galaxies are more abundant at $z>2$ than what was expected from {\it HST} surveys. \citet{Suess22} find that star-forming galaxies are more concentrated in terms of their stellar mass than in their rest-frame optical light at $1.0 < z < 2.5$ compared to previous estimates using {\it HST} data. \citet{Guo23} find galaxy disks hosting stellar bars at redshifts as high as $z\approx2.3$, suggesting that galaxy disks like those in the local Universe may have been in place much earlier than anticipated. Recently, \citet{Pandya23} find that the majority of low-mass galaxies at $z>1$ are prolate, cigar-shaped systems instead of oblate disks. 

This work examines the size-mass relation of star-forming galaxies and the relationship between star-formation activity and S\'{e}rsic index out to $z=5.5$. We use {\it JWST} data from CEERS, spanning an area of 97 square arcmin, as well as {\it HST} data from the Cosmic Assembly Near-Infrared Deep Extragalactic Legacy Survey (CANDELS; \citealt{Grogin11, Koekemoer11}), to select star-forming and quiescent galaxies at stellar masses greater than $10^{9.5} \ \textrm{M}_{\odot}$. We use 13 bandpasses to accurately and precisely constrain the effective radii and S\'{e}rsic indices at rest-frame 5,000\AA \ of more than 2,000 galaxies to $z=5.5$. 

This paper is structured as follows. In Section \ref{sec:Data}, we describe the data we use in this paper. In Section \ref{sec:sample}, we list the cuts we apply to select our sample of galaxies. Section \ref{sec:methods} describes our measurements of galaxy morphologies and the associated uncertainties, classification of galaxies into star-forming and quiescent systems, and how the size-mass relation is estimated. The analysis of the size-mass relation and the morphology-quenching relation is described in Section \ref{sec:results}. We interpret our results in Section \ref{sec:discussion}. We conclude in Section \ref{sec:conclusions}. 

In this paper, we assume a $\Lambda$CDM cosmology with $\Omega_{\Lambda}=0.3$, $\Omega_M = 0.7$, and $H_0=70 \ \textrm{km s}^{-1} \ \textrm{Mpc}^{-1}$. All magnitudes are in the AB photometric system \citep{Oke83}. 

\section{Data} \label{sec:Data}
\subsection{CANDELS and CEERS Imaging}

The CEERS Team has provided high quality mosaics as a part of the {\it JWST} Early Release Science (ERS) program \citep{Finkesltein22}. The CEERS mosaics cover 97 sq. arcmin and overlap with the Extended Groth Strip (EGS), a {\it Hubble Space Telscope (HST)} legacy field. Mosaics in 13 bandpasses are used in this work: F606W and F814W (ACS); F105W, F125W, F140W, and F160W (WFC3); and F115W, F150W, F200W, F277W, F356W, F410M, and F444W (NIRCam). We include {\it HST} data from CANDELS \citep{Grogin11, Koekemoer11} so that the rest-frame $5,000$\AA \ is sampled at the lowest redshifts considered in this work. The 5$\sigma$ depths in each filter are correspondingly 28.62 and 28.30 (ACS); 27.11, 27.31, 26.67, and 27.37 (WFC3); 29.15, 29.00, 29.17, 29.19, 29.17, 28.38, and 28.58 mag (NIRCam). All mosaics have a pixel scale of 30 milli-arcseconds. The CEERS survey observed the EGS field in 10 NIRCam pointings. Objects are detected in an error-weighted mosaic of the F277W and F356W bandpasses.

Data reduction was performed by \citet{Bagley23}. In short, JWST Calibration pipeline v1.7.2 and CRDS pmap 0989 were used to process the pointings from June 2022, whereas JWST Calibration pipeline v1.8.5 and CRDS pmap 1023 were used to process pointings from December 2022. For all pointings, the images were reduced using a custom pipeline developed by the CEERS team, which includes $1/f$ noise subtraction and snowball removal. Lastly, the mosaics were aligned using astrometry from Gaia-EDR3 \citep{Gaia21}, mapped to a pixel scale of 0.03"/px, and finally background subtracted.

We adopt simulated NIRCam point spread functions (PSFs) generated using WebbPSF \citep{Perrin14}. The PSF full-width at half-maximum (FWHM) of the {\it JWST} NIRCam bandpasses used here ranges from 0.040" to 0.145"\footnote{\url{https://jwst-docs.stsci.edu/jwst-near-infrared-camera/nircam-performance/nircam-point-spread-functions}}. The HST ACS and WFC3 bandpasses used in this work have PSF FWHM ranges from 0.10" to 0.19" \cite{Skelton14}.  

\subsection{Photometric Redshifts, Rest-Frame Colors, and Stellar Masses}
\label{sec:sed_fitting}

Photometric redshifts and rest-frame colors for our sample are estimated using {\sc eazy} \citep{Brammer08} and adopted from the fits performed by \citet{Barro23}. {\sc eazy} measures photometric redshifts by fitting non-negative linear combinations of templates to photometric observations and derives probability distribution functions of the redshift. Templates are supplied by users. For our sample, we adopt the standard ``tweak\_fsps\_QSF\_12\_v3'' templates, which are made of 12 flexible stellar population synthesis ({\tt FSPS}, \citealt{Conroy10}) templates, as recommended by the {\sc eazy} documentation. A prior on the fluxes of galaxies in F277W is assumed to assign low probabilities at very low redshifts and for bright galaxies at very high redshifts. 

{\sc eazy} is run on a flux-limited subset of the CEERS catalog (see Section \ref{sec:sample}). To assess the accuracy of the photometric redshifts inferred by {\sc eazy}, we compare these against spectroscopic redshifts of 1,187 galaxies in the All-Wavelength Extended Groth Strip International Survey (AEGIS; \citealt{Davis07}) that were compiled by \citet[][see references therein]{Stefanon17}. We use the bias and scatter scatter using the normalized median absolute deviation (NMAD) to quantify the accuracy of the photometric redshifts \citep{Dahlen13}. We find a bias of -0.01, such that the photometric redshifts very slightly overestimate the spectroscopic redshifts, and $\sigma_{\textrm{NMAD}} = 0.02$. 

Stellar masses for the galaxies in our sample are estimated using {\sc fast} \citep{Kriek09} and are also adopted from \citet{Barro23}. Exponentially declining star-formation histories are assumed and the minimum star-formation history timescale is set to $\log \tau \ (\textrm{yr}) = 8.5$. The metallicity is fixed to solar and a \citet{Calzetti00} attenuation curve is assumed. The \citet{BC03} stellar population synthesis templates are assumed and the initial mass function is set to that by \citet{Chabrier03}. 

\section{Sample Selection} \label{sec:sample}

\begin{figure}[t!]
\includegraphics[width=0.45\textwidth]{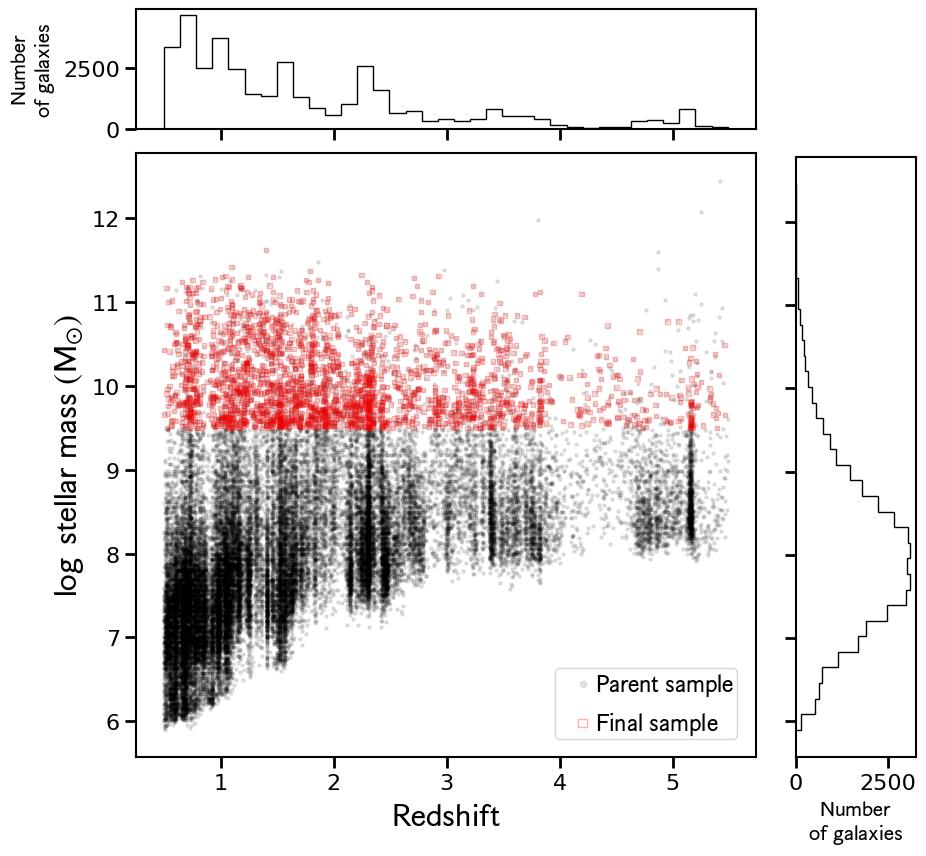}
\caption{The distribution of the parent and final samples (see Section \ref{sec:sample} in terms of stellar mass and redshift and over the redshift range examined in this paper, $0.5 < z < 5.5$. The parent sample is displayed as black dots and the final sample is shown as empty red squares. The redshift and stellar mass distributions of the parent sample are presented as histograms in the top panel and right panels, respectively. We estimate a stellar mass completeness limit of $\sim10^8 \ \textrm{M}_{\odot}$.}
\label{fig:sample_completeness}
\end{figure}

% start with 101808
% Barro cat selects sources with F356W AB mag < 28.5 --> 53885 sources
% match to Nandra+CEERS+Barro --> 91 matches
% cuts in RH_f200w-F200W mag parameter space used to select point sources:
% -0.001 * F200W_mag + 0.075 AND -0.001 * F200W_mag + 0.06 AND f200W_mag < 28
% 

We start with all sources in the CEERS photometry catalog (v0.51.2). There is a total of 101,808 sources. We first select sources that are brighter than 28.5 mag in F356W. This cut removes 47,923 objects and leaves 53,885. These 53,885 galaxies are the parent sample from which we shall produce our final sample. The parent sample is shown in Figure \ref{fig:sample_completeness} as black dots. 

%To ensure high signal-to-noise ratios (S/N) in the photometry, we select galaxies with F356W magnitude brighter than 28.5. This cut removes low-mass (often with stellar mass $< 10^9 \ \textrm{M}_{\odot}$) and high-redshift galaxies (mostly at $z>3$) and reduces the sample by 41 objects with 15,491 remaining. 

%We remove unresolved point sources and stars by choosing objects with stellarity index CLASS\_STAR $< 0.1$. CLASS\_STAR is measured by Source Extractor and comes from the 3D-HST catalog. Applying this cut removes 3,919 sources and leaves 11,613 galaxies. 

\begin{figure*}[t!]
\includegraphics[width=\textwidth]{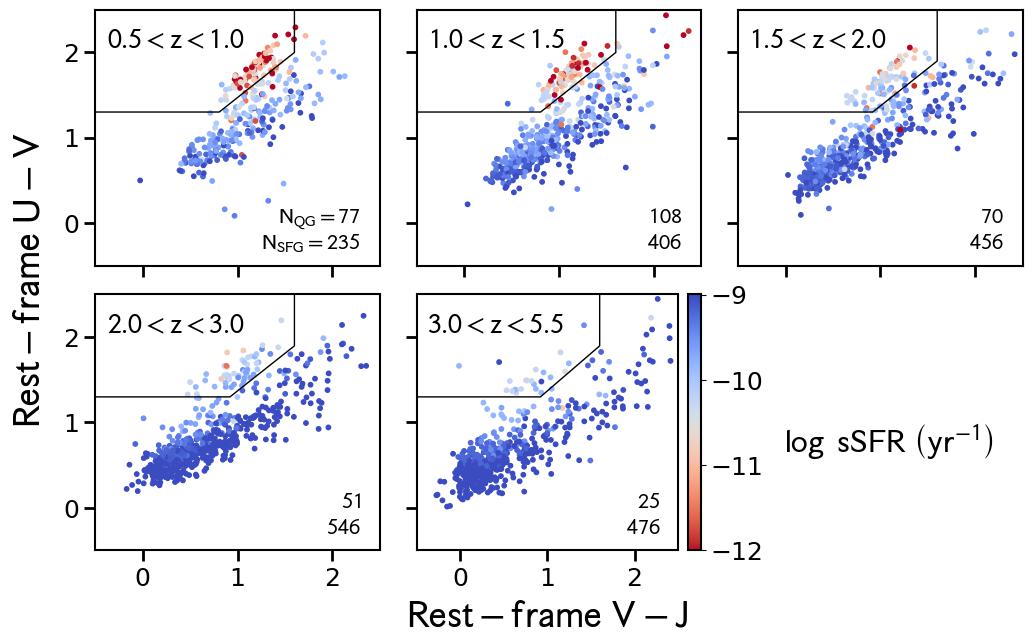}
\caption{Rest-frame {\it U-V} and {\it V-J} color distributions of the sample examined in this paper. We split our sample into five redshift bins, which are indicated in the top left corner of each panel. Galaxies are colored by sSFR, which is measured by SED fitting (see Section \ref{sec:sample}). Quiescent galaxies are indicated by red points. Color selection criteria for quiescent galaxies developed by \citet{Williams09} are indicated by the thin lines. Galaxies are considered quiescent if they lie within these lines. The selection criteria by \citet{Williams09} recover well the color-color distributions of quiescent galaxies out to $z\sim3$. The numbers of quiescent and star-forming galaxies in each redshift bin are shown in the bottom right corner of each panel.  }
\label{fig:sample}
\end{figure*}

We remove unresolved point sources and stars by following the procedure as described in Section 3 by \citet{Holwerda23}. Unresolved objects are identified by comparing the half-light radius in F200W, measured by Source Extractor, against the F200W magnitude. Stars and unresolved objects lie on a tight sequence of compact objects that extends to bright magnitudes. We denote the stellar locus in this parameter space using 
$$\left| R_{h, F200W} + 0.001 m_{F200W} + 0.065\right| < 0.01,$$
where $R_{h, F200W}$ is the half-light radius in arcsec and $m_{F200W}$ is AB magnitude. The inequality above applies only to objects brighter than 28th magnitude in F200W. There are 572 objects that are removed as they lie on the stellar locus, which leaves 53,313 galaxies remaining. 

The main goal of this work is to determine the evolution with redshift of the rest-frame optical size-mass relation of star-forming galaxies. Sizes are measured in the rest-frame $5,000$\AA \ to be consistent with previous studies. To ensure that there is at least one bandpass at shorter and longer wavelengths than the rest-frame $5,000$\AA \ for each galaxy, we select galaxies at redshifts $0.5 \leq z \leq 5.5$. This cut excludes 15,721 galaxies with 37,592 remaining. 

This work relies on the morphological measurements of galaxies spanning a wide range in redshift. To ensure high S/N in as many bandpasses as possible, we select galaxies with stellar mass $> 10^{9.5} \ \textrm{M}_{\odot}$. Over the redshift range examined in this work, we estimate the completeness limit in mass to be $\sim10^8 \ \textrm{M}_{\odot}$, similar to what \citet{HuertasCompany23} found. This limit can be seen as the peak in the histogram in the right side panel of Figure \ref{fig:sample_completeness}. This cut in stellar mass is well above our estimated completeness limit and reduces the sample by 34,798 and leaves 2,794 galaxies. 

%Galaxies are further selected to have spectroscopic redshifts or photometric redshifts with small uncertainties. Photometric redshifts and physical properties of galaxies are derived by spectral energy distribution (SED) fitting. We adopt the photometric redshift and physical property catalog of CEERS galaxies derived using the Dense Basis SED-fitting technique \citep[][K. Iyer, priv. communication]{Iyer19}. Of the 11,613 left, 408 have spectroscopic redshifts. The rest have only photometric redshifts. Photometric redshifts are estimated via a Bayesian technique and thus have posterior distributions. For galaxies with spectrosopic redshifts, the photometric and spectroscopic redshifts are in good agreement: the typical discrepancy between the photometric and spectroscopic redshifts is $\frac{ z_{\rm{phot}} - z_{\rm{spec}} }{1 + z_{\rm{spec}} } = 0.03$. We use the difference between the 16th and 84th percentiles of each photometric redshift posterior distribution, which we define as the photometric redshift posterior width, to quantify the redshift uncertainty. We select galaxies with photometric redshift posterior width $< 0.5$. This cut removes 5,453 galaxies with 6,160 remaining. 

We remove galaxies hosting active galactic nuclei (AGN), which may not be fitted well with one-component S\'{e}rsic models, as follows. To remove X-ray-detected AGN, we match our catalog to the {\it Chandra} X-ray catalog produced by \citet{Nandra15}. These observations reach a typical depth of 800 ks across the EGS field. We use a search radius of 0.5 arcsec and find 91 matches to the \citet{Nandra15} catalog, 74 of which are in our mass-selected sample above. All 74 of these matches are removed from our sample, leaving 2,720 galaxies. 

Early {\it JWST} surveys revealed a new population of compact objects with blue SEDs at short wavelengths and very red SEDs at long wavelengths \citep{Barro23, Labbe23, Matthee23}. The colors of such objects can be explained by combining dust-reddened quasar light with bluer light from young stellar populations. Due to the puzzling nature of such objects, we apply a strict color cut of F277W - F444W $>$ 1.5, which we adopt from \citet{Barro23}, to remove these objects. This cut excludes 11 galaxies and 2,709 remain. 

Since this work aims to simultaneously fit {\it HST} and {\it JWST} images of a sample of galaxies, we remove galaxies that are not detected in {\it HST} images. There are 50 such galaxies removed from the sample, which leaves 2,659. 

We visually inspected these 2,659 galaxies for bright neighboring galaxies or stars that could adversely affect the 2D S\'{e}rsic fits, galaxies lying on the edge of the mosaic, and spurious sources that were identified by the hot Source Extractor run as described in Section \ref{sec:sample}. We remove 132 galaxies: 35 are removed due to having bright nearby stars; 77 are removed due to their proximity to the edge of the mosaic; and 20 are removed due to being spurious sources. After excising these sources, 2,527 galaxies remain. 

%below is from tweaking Ormerod
To remove objects that have poor fits with large residuals, we use a cut in the residual flux fraction (RFF, see Section \ref{sec:RFF}). A large RFF signifies that much of the light in the galaxy is not adequately modeled, and hence morphological parameters may be inaccurate (e.g., bright spiral arms and bars that are not accounted for in 2D S\'{e}rsic fits). We reject objects with $\left| RFF \right| > 0.5$ which excludes 15 galaxies due to large residuals, leaving 2,512 galaxies. We also exclude 62 galaxies that have S\'{e}rsic indices that lie beyond the bounds we set for our fits ($0.3 < n < 8$) or effective radii that are too small or too large (sizes must lie between 0.009'' and 3''), with 2,450 remaining.

%We exclude AGN detected in the mid-infrared by applying the color-color selection criteria developed by \citet{Donley12} to galaxies detected in all four {\it Spitzer} IRAC bandpasses in the matched 3D-HST catalog. There are 22 galaxies satisfying these criteria, two of which survived the previous cuts we applied. We thus exclude these two, leaving 2,339 galaxies. 

%Lastly, we visually inspected the remaining 1,076 galaxies for bright neighbors that adversely affect the 2D S\'{e}rsic fits and galaxies lying on the edge of the mosaic in the bluest (F606W) or reddest (F444W) bandpasses. We find two galaxies that meet the first condition and six that meet the second, thus eight additional galaxies are removed. 

The final sample contains 2,450 galaxies. These are shown as red empty squares in Figure \ref{fig:sample_completeness} in terms of stellar mass and redshift. In Figure \ref{fig:sample}, the final sample is displayed in rest-frame {\it UVJ} color-color diagrams and in five redshift bins. The galaxies in our sample populate the regions of color-color space spanned by star-forming and quiescent galaxies (see Section \ref{sec:uvj}). 

\section{Methods} \label{sec:methods}

\subsection{Two-dimensional S\'{e}rsic Model Fitting}

\begin{figure}
    \includegraphics[width=0.45\textwidth]{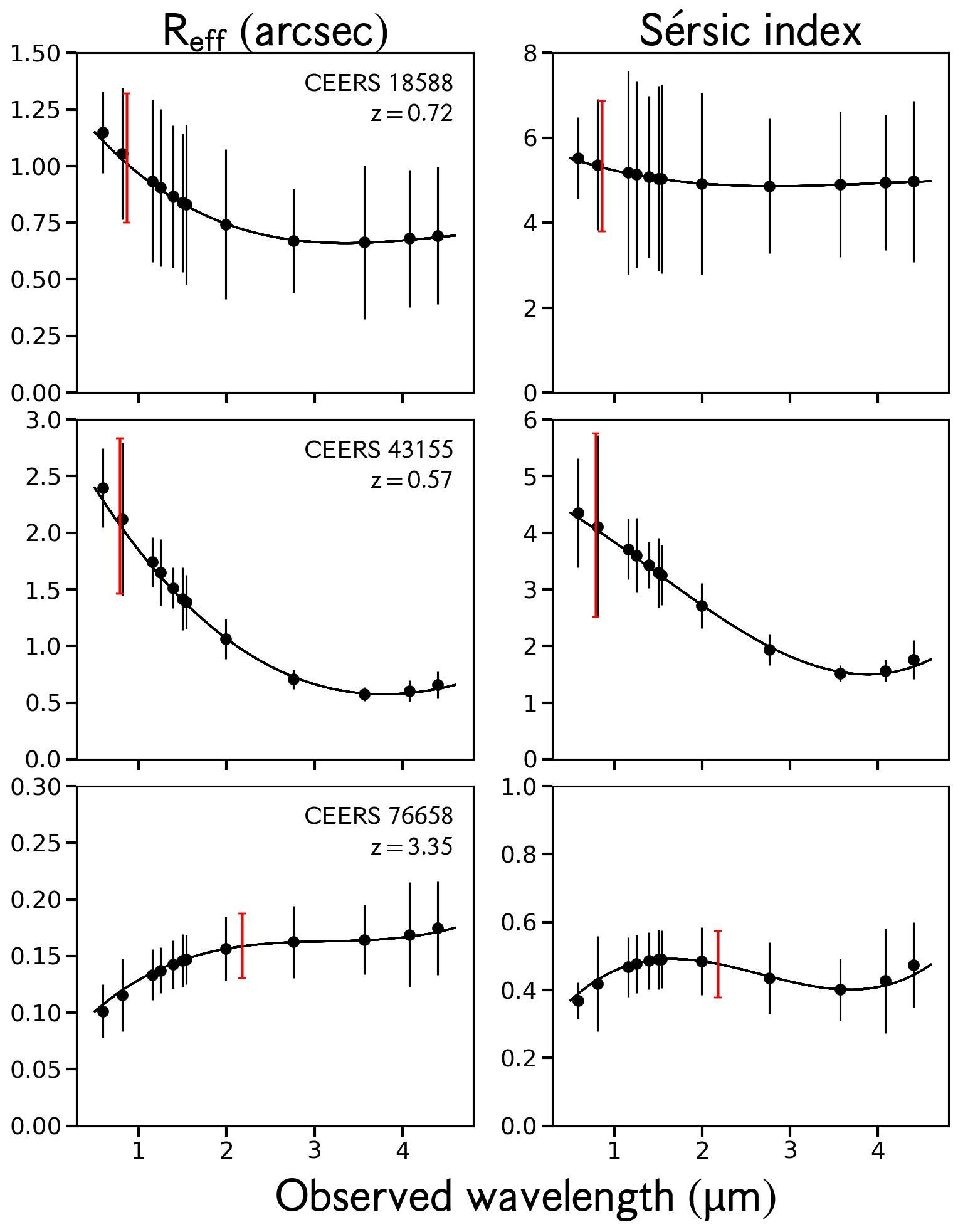}
    \caption{Multi-wavelength fits to three example galaxies. Fits for a different galaxy are shown in each row. Measurements in each bandpass from the fits are shown as dots with error bars. Effective radii and S\'{e}rsic indices in different bandpasses are shown in the left and right columns, respectively. The third-order fits to the images are shown as solid lines. Red error bars indicate parameters in the rest-frame $5,000$\AA \ derived from the multi-band fits.}
    \label{fig:example_fits_reff_n}
\end{figure}

We use the {\sc galfitm} tool \citep{Barden12, Haussler13, Vika13, Haussler22} to perform simultaneous multi-wavelength morphological fits of galaxies in our sample. {\sc galfitm} is used as it has been shown that morphological fits using this tool result in more stable and accurately measured parameters \citep{Haussler13}. {\sc galfitm} is built upon the {\sc galfit} code \citep{Peng02, Peng10}, which performs morphological fits to images in individual bandpasses. Morphologies are quantified with two-dimensional (2D) S\'{e}rsic functions \citep{Sersic68}, which are then convolved with the PSF at different wavelengths. 

We construct error maps for each bandpass, which take into account Poisson noise from galaxies, instrumental noise, and errors due to background subtraction. For the NIRCam bandpasses, these error maps have been produced by the CEERS team \citep[see][for more details]{Bagley23}. For the {\it HST} bandpasses, we estimate these maps following the procedure as described in Section 4.1 by \citet{vdw12}, which makes use of exposure time mosaics and weight maps. 

The images used to perform the fits are cutouts from the {\it HST} and {\it JWST} mosaics. Each cutout is square and has side length equal to 20 times the half-light radius measured by Source Extractor in the F200W bandpass. For each galaxy in our sample, we also fit neighboring galaxies that are at least half as bright as the target galaxy in F356W, which is the redder bandpass used to detect galaxies in CEERS (see Section \ref{sec:Data}). Neighboring galaxies are fitted provided they lie within the extent of the cutout of the target. Fainter objects are masked out using segmentation maps determined by Source Extractor.

\begin{figure*}
    \includegraphics[scale=0.34]{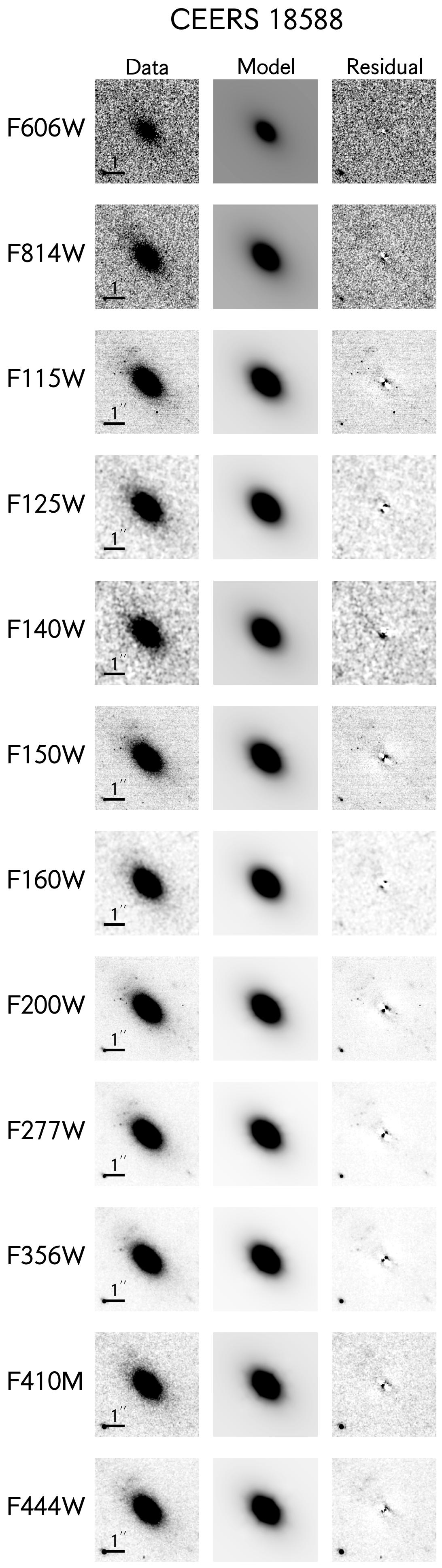}
    \hspace{0.35cm}
    \includegraphics[scale=0.34]{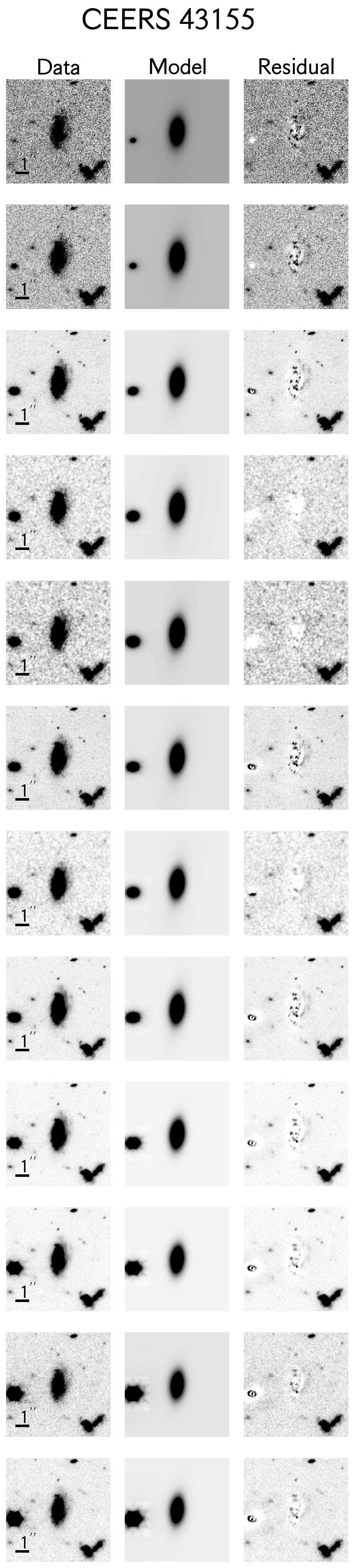}
    \hspace{0.35cm}
    \includegraphics[scale=0.34]{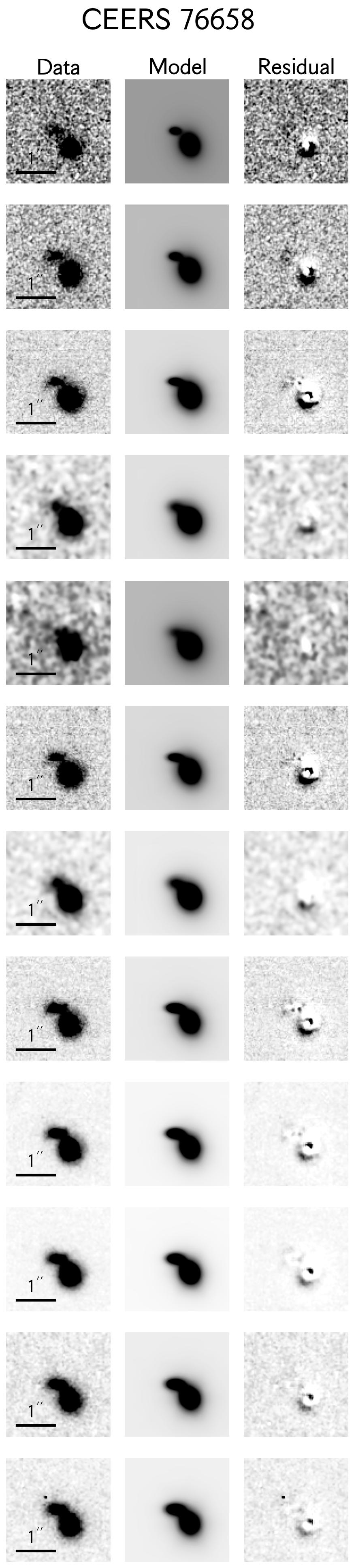}
    \caption{Multi-wavelength fits to the same three example galaxies shown in Figure \ref{fig:example_fits_reff_n}. In each set of three columns, the left column displays postage stamps in all 12 bandpasses in which each galaxy is observed. The black bars in the lower left corner of each set of plots indicate an angular scale of 1 arcsec. None of these galaxies were observed in F105W, thus we do not show the fits in that bandpass. In the middle columns are the best-fitting models in each bandpass derived from the simultaneous multi-band fits. In the rightmost columns are images of the residuals, i.e., the models subtracted from the data. Our fits perform well in general, and bright neighbors are fit well, as can be seen in the fits for CEERS 43155 and 76658.}
    \label{fig:example_fits_imgs}
\end{figure*}

For our simultaneous multi-wavelength fits, we largely follow what is commonly done in the literature, albeit with some modifications. Earlier studies that performed multi-wavelength fits use up to seven bandpasses in their fits and often assume second-order polynomials to model both the effective radius and S\'{e}rsic index as a function of wavelength \citep{Haussler13, Vika13, Nedkova21}. This is done as galaxies are observed to vary in size and compactness as a function of wavelength, due to, e.g., the presence of a bulge or disk \citep[e.g.,][]{Vulcani14}.  

After trial and error, we instead assume third-order polynomials to model both the effective radius and S\'{e}rsic index as a function of wavelength. For many galaxies, we find that their effective radii and S\'{e}rsic indices, when plotted as a function of wavelength, present at least one inflection point. Examples of multi-wavelength fits are shown for three galaxies as points with error bars in Figure \ref{fig:example_fits_reff_n}. In particular, for CEERS 43155, displayed in the middle row, its S\'{e}rsic index as a function of wavelength decreases steeply from $\sim0.6$ to $\sim3 \mu \textrm{m}$ and is nearly constant at longer wavelengths. Modeling this behavior is difficult if second-order polynomials are assumed, whereas third-order polynomials have enough degrees of freedom to fit data like these. 

Following \citet{Nedkova21}, the axis ratio, central coordinates, and position angle are each fitted to constant values that do not depend on wavelength. 
%are left to vary but are held constant across all wavelengths. 
Following \citet{Haussler13}, we also assume Chebyshev polynomials of the first kind \citep{AbramowitzStegun65} in the multi-wavelength fits. The magnitudes of the models as a function of wavelength are left to vary freely in all 13 bandpasses. There are thus 25 free parameters in total for each multi-wavelength fit. For all galaxies, we restrict the effective radius in each bandpass to lie between 0.3 and 100 pixels (0.009 and 3 arcsec) and the S\'{e}rsic index to lie between 0.3 and 8. 

%To perform the multi-wavelength fits, fourth-order polynomials are fitted to the effective radii and S\'{e}rsic indices as a function of wavelength, which are derived from the single-band fits. The fitted polynomials serve as initial guesses for the multi-wavelength fits. The background in each bandpass from the single-band fits is assumed to be the same in each bandpass for the multi-wavelength fits. Surface brightness at the effective radius in each bandpass is allowed to vary in the multi-wavelength fits. 

Examples of multi-wavelength fits are shown in Figure \ref{fig:example_fits_imgs} for the same galaxies whose fitted parameters are shown in Figure \ref{fig:example_fits_reff_n}. For each galaxy, there are three columns. The left column shows the galaxy in each bandpass from top to bottom in order of increasing wavelength. The middle column shows the best-fitting model from the multi-wavelength fit. The right column shows the model subtracted from the galaxy image, i.e., the residual image. In general, our fits perform well. One galaxy, CEERS 76658, is fitted well despite a bright and close neighbor. This suggests that our masking technique results in reasonable fits. For most fits, residuals of the order of 10\% remain, but these are expected, given that one-component S\'{e}rsic models are likely not the most appropriate description of the morphologies of most galaxies \citep{Peng02}. In Appendix \ref{sec: apdx}, we compare our multi-wavelength fits to fits that are performed bandpass-by-bandpass by McGrath et al. (in prep.) using {\sc galfit}. We find good agreement between their fits and ours. 

\subsection{Residual Flux Fraction}
\label{sec:RFF}

%using template from Ormerod then tweaking
%may be best to cite Pagul+21 HFF paper, explains this well and already cosigned by BM
To identify galaxies with poor morphological fits and remove them from our sample, we use the residual flux fraction (RFF; see Section \ref{sec:sample}). The RFF is a measure of the signal in the residual image that cannot be attributed to fluctuations in background sky values \citep{Hoyos12}. Following \citet{Margalef-Bentabol16}, we define the RFF as:

$$ RFF = \frac{\sum\limits_{j,k \in A}  \left| I_{j,k} - I_{j,k}^{GALFITM} \right| - 0.8 \sum\limits_{j,k \in A} \sigma_{B_{j,k}}}{FLUX\textunderscore AUTO}$$

\noindent where $I_{j,k}$ is the flux in the $j$th and $k$th pixels in the science image, $I^{GALFITM}$ is the model image estimated by {\sc galfitm}, $\sigma_{B}$ is the background error due to sky subtraction, and FLUX\textunderscore AUTO is the flux of the galaxy calculated by Source Extractor. These are all in the bandpass that is closest to the rest-frame 5,000\AA \ of the galaxy.

%"The factor of 0.8 in the numerator ensures that the expected value of the RFF is 0 for a Gaussian noise error image (Hoyos et al. 2011). 

The factor of 0.8 in the numerator ensures that the average value of the RFF vanishes to zero when a Gaussian error image with constant variance is encountered \citep{Hoyos11}. We calculate the RFF within an elliptical aperture of radius equal to 1.5 times the Kron radius, where we define the Kron radius as the semi-major axis of the Kron ellipse. Calculating the RFF over a large radius leads to the RFF decaying to zero, where the outer areas can dominate the calculation, even if there is a complex residual at the centre of the image. Any neighboring sources within the Kron aperture are masked. 

%the \textcolor{red}{1.5 Kron radii of the galaxy}, where we define the Kron radius as the semi-major axis of the Kron ellipse. Calculating the RFF over a large radius leads to the RFF decaying to zero, where the outer areas can dominate the calculation, even if there is a complex residual at the centre of the image.

We calculate the background sigma term following \citet{Margalef-Bentabol16}:

$$ \sum\limits_{j,k \in A} \sigma_{B_{j,k}} = N \langle \sigma_{B} \rangle$$

\noindent where $\langle \sigma_{B} \rangle$ is the mean value of the sky background. This is estimated by computing the mean flux estimated by placing empty apertures in random locations of the background-subtracted science image. The value of $N$ is the number of pixels within the Kron aperture that we describe above.

%the background sigma (sky value) for the whole image. Following \citet{Ormerod23}, this is calculated by placing apertures on blank areas of sky in the science image, and calculating the mean value of these regions. The value of $N$ is the number of pixels within an aperture of 1.5 times the Kron radii of the galaxy that we are using for the RFF calculation."

%"We find many faint sources with RFF < 0. We interpret this as overfitting: RFF < 0 means that there is less noise in the image than expected, so, the code has modelled the noise away. For larger galaxies this problem is of course less severe, as the code has to model more pixels with higher signal-to-noise, so it has less freedom."-Hoyos+11

\subsection{Morphological Parameter Uncertainties}

We follow the technique developed by \citet{vdw12}, albeit with minor changes, to compute uncertainties in individual bandpasses on the effective radii and S\'{e}rsic indices derived from the multi-wavelength fits. Uncertainties on rest-frame quantities are obtained by taking relative errors on the parameters in the bandpass that is closest to the rest-frame $5,000$\AA \ at the redshift of each galaxy in our sample. 

%For the multi-band fits, uncertainties are obtained by taking relative errors from the single-band fits and multiplying them by the values inferred from the multi-band fits. 

Uncertainties on the effective radii and S\'{e}rsic indices in individual bandpasses are derived as follows. Following Sec. 5.1 by \citet{vdw12}, for each galaxy in our sample, we first search for galaxies with derived properties most similar to the target galaxy in a given bandpass. This is done by computing the Euclidean distance 
$$p_{i,j} = $$
$$\sqrt{\left(\frac{m_i - m_j}{\sigma m}\right)^2 + \left(\frac{\log n_i - \log n_j}{\sigma (\log n)}\right)^2 + \left(\frac{\log r_i - \log r_j}{\sigma (\log r)}\right)^2},$$
where $m_i, n_i$, and $r_i$ are the magnitude, S\'{e}rsic index, and effective radius of the target galaxy, and parameters with subscript $j$ denote parameters pertaining to another galaxy in the sample. In the equation above, $\sigma$ denotes the standard deviation of the respective parameter. Hence, these distances are dimensionless and normalized relative to the distribution of parameters in the sample. 

For the 11 galaxies most similar to the target galaxy, i.e., the 11 galaxies with the smallest $p_{i,j}$ as defined above, we do the following. First, we compute the differences in each parameter between these 11 galaxies and the target galaxy. Then, we compute the difference between the 16th and 84th percentiles of the distribution of differences defined above. This difference between the 16th and 84th percentiles, $q_{i,j}$ ($\delta_{i,j}$ in \citealt{vdw12}), is computed for each galaxy in our sample. We account for the S/N of the target by multiplying $q_{i,j}$ by the S/N of the target. 

%$$ q_{i,j} = q_{84} - q_{16} \left( r_j - r_i \right) $$

We determine uncertainties on the parameters as follows. First, we find the 25 nearest galaxies\footnote{Altering this number by a factor of two does not significantly affect our results.} to a target galaxy using $p_{i,j}$ defined above. For these 25 nearest galaxies, we find the average $q_{i,j}$, as described earlier, and divide by the S/N of the target.  

For the computations described above, we keep the effective radii and S\'{e}rsic indices in logarithmic units and leave magnitudes as they are. The final uncertainties on the parameters are converted to linear units, e.g., $\delta r = \ln(10) r \delta \log r$. This procedure yields relative errors ($\lesssim15\%$) that are consistent with what is found in Table 2 by \citet{vdw12} and Table A1 by \citet{Nedkova21} for galaxies with F160W magnitudes $<26$, as is appropriate for our sample. 

The minor differences between the procedure above and what \citet{vdw12} do are as follows. \citet{vdw12} have a parent sample that is $\approx2.5$ times larger than ours (6,492 vs. 2,450). These authors select the 200 closest galaxies to the target prior to computing $q_{i,j}$, whereas we select the 11 closest. We downscale this number of 200 to 11 to make it more appropriate for our sample as follows. We first scale down this number by taking the ratio of the size of the sample used by \citet{vdw12} and ours (6492 / 2450). We then take the cube of this ratio as the Euclidean distance $p_{i,j}$ is calculated in three dimensions. \citet{vdw12} compute $q_{i,j}$ using the differences between parameters inferred from deep and shallow imaging in CANDELS, whereas we calculate $q_{i,j}$ using the differences between the parameters of the 11 closest galaxies to the target and those of the target itself. In essence, our uncertainties are computed by estimating the ensemble properties of galaxies that are similar to a target galaxy. 

%we compute the standard deviation of the effective radius and S\'{e}rsic index. These quantities serve as the uncertainties on these parameters. This process is performed in each of the 13 bandpasses of our dataset. Typical relative uncertainties in the effective radius and S\'{e}rsic index are $\sim17$\% for both parameters, similar to what \citet{vdw12} obtained. Examples of uncertainties in the single-band fits are shown as black error bars in Figure \ref{fig:example_fits_reff_n}. 

For uncertainties on the rest-frame parameters, we assume the relative uncertainties in the bandpasses closest to the rest-frame 5,000\AA \ at the redshifts of the galaxies in our sample. To obtain the total uncertainties on the rest-frame parameters, we multiply these relative uncertainties by the parameters inferred from the mutli-wavelength fits at the wavelengths that correspond to the rest-frame 5000\AA. This approach is similar to what \citet{Nedkova21} do, who instead compute uncertainties in rest-frame optical quantities by interpolating between the uncertainties on the parameters in the bandpasses that surround the rest-frame 5,000\AA. Parameters and their respective uncertainties in the rest-frame 5000\AA \ are shown as red vertical error bars in Figure \ref{fig:example_fits_reff_n}.

%derived from the multi-wavelength fits, we assume the relative uncertainties from the bandpasses closest to the rest-frame 5000\AA \ at the redshifts of the galaxies in our sample. Thus, to obtain the total uncertainties on the rest-frame parameters, we multiply these relative uncertainties by the parameters inferred from the mutli-wavelength fits. Uncertainties on the effective radius and S\'{e}rsic index in the rest-frame 5000\AA \ are typically 2\% for both parameters. 

\iffalse
\begin{itemize}
    \item We follow the technique of \citet{vdw12}, albeit with minor changes, to compute uncertainties on the size and Sersic index. 
    \item For each galaxy in our sample, we first compute a 3D Cartesian distance of the difference between the normalized parameters of the galaxy and those of all other galaxies in the sample. The uncertainties on the size and Sersic index are the standard deviations of the sizes and Sersic indices of the 30 ``nearest'' galaxies according to the distance defined above. We do this for each bandpass. Relative uncertainties for each parameter vary with wavelength, but typical size and Sersic index uncertainties are $\sim15$\% and $\sim20$\%, respectively. These are similar to what \cite{vdw12} measure. 
\end{itemize}
\fi

\subsection{Separating Star-forming and Quiescent Galaxies using the Rest-frame {\it UVJ} Color-color Diagram}
\label{sec:uvj}

Galaxies are often separated into star-forming and quiescent using the rest-frame {\it U-V} and {\it V-J} colors \citep[e.g.,][]{Williams09, Whitaker11}. These colors are often interpreted as indicators of the age and dust content of a stellar population, respectively. We adopt the redshift-dependent color-color criteria developed by \citet{Williams09} to identify quiescent galaxies. These criteria are motivated by the observed distribution of galaxies in color-color space as a function of redshift. 

The rest-frame {\it UVJ} colors of galaxies in our sample are shown in Figure \ref{fig:sample}. Galaxies are presented as dots colored by log(sSFR), with blue and red colors indicating high and low sSFR, respectively in Figure \ref{fig:sample}. The selection criteria by \citet{Williams09}, shown as thin dark lines, separate quiescent and star-forming galaxies, which correspond to red and blue dots, respectively. Despite the few quiescent galaxies observed at $3.0 < z < 5.5$, galaxies with the lowest sSFRs at these redshifts tend to lie within the selection region. 

To check the nature of these putative quiescent galaxies, we determine which galaxies lie in the observed F150W-F277W, F277W-F444W color-color selection wedge developed by \citet{Long23}. These authors determined where in this observed color-color space quiescent galaxies at $3 < z < 6$ lie. We find that 20 of the 25 galaxies in the quiescent region in the {\it UVJ} diagram at $3.0 < z < 5.5$ lie in the \citet{Long23} wedge. This suggests that, at least for the sample of galaxies examined in this work, our selection criteria perform well over the entire redshift range we examine.

\subsection{Measuring the Size-mass Relation for Star-forming Galaxies}
\label{sec:fit_size_mass}

The goal of this work is to determine the evolution of the relationship between effective radius and stellar mass with redshift for star-forming galaxies out to $z=5.5$. Star-forming galaxies are selected for these measurements as quiescent galaxies become increasingly rare with increasing redshift. The cuts we have applied result in too few quiescent galaxies in each redshift bin to adequately measure the mass-size relation for quiescent galaxies.

\begin{figure*}[t!]
\includegraphics[width=\textwidth]{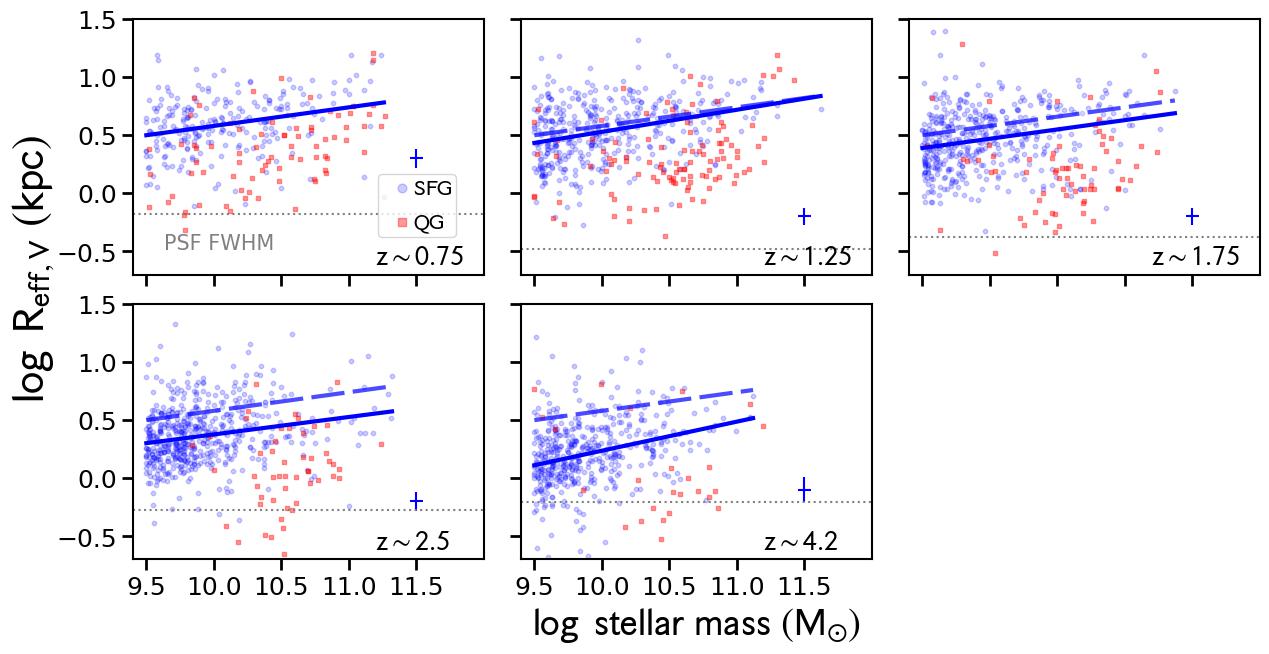}
\caption{The relation between the size at rest-frame 5,000\AA\ and stellar mass of galaxies in our sample as a function of redshift. Star-forming galaxies and quiescent galaxies are shown as small blue dots and small red squares, respectively. Redshift increases from left to right. The median redshift in each bin is indicated in the lower right corner of each panel. The mean uncertainties in size and stellar mass are indicated by the error bars in the lower right corner of each panel. The best-fit size-mass relations are plotted as solid blue lines. The dashed blue line is the best-fitting relation at $z \sim 0.7$ plotted in the other bins for comparison. We find that the slope is roughly constant out to $z \sim 2.5$ and shows a minor increase at $z \sim 4.2$. The PSF FWHM of the bandpass that is closest to $5,000$\AA\ in the rest frame is shown in dotted lines in each panel. Only a handful of galaxies in our sample have sizes that are comparable to the PSF FHWM in each redshift bin.}
\label{fig:size_mass}
\end{figure*}

Following earlier studies \citep{vdw14, Dimauro19, Mowla19, Nedkova21}, we parameterize the size-mass relation by assuming a linear relation between the logarithm of the stellar mass and the logarithm of the effective radius in units of kpc and in the rest-frame 5,000\AA, or more simply, a power law. The parameters of this linear relation are the intercept at a stellar mass of $5\times10^{10} \ \textrm{M}_{\odot}$ and the slope. Following \citet{vdw14}, we can also measure the intrinsic scatter of the size-mass relation under the assumption that it behaves like a log-normal distribution (i.e., assume Gaussian scatter when fitting the log of the mass and size). 

We fit for these three parameters (intercept A, slope $\alpha$, and intrinsic scatter $\sigma_{\log \ \textrm{Reff}}$) using a Markov Chain Monte Carlo (MCMC) approach following \citet{Nedkova21}. We use the {\tt emcee} tool \citep{ForemanMackey13} to estimate the posterior distributions for each parameter. Priors are assumed to be uniform with the range on each prior set to [-1.0, 1.5], [0.0, 2.0], and [0.0, 2.0] on log A, $\alpha$, and $\sigma \left(\log \ \textrm{R}_{\textrm{eff}}\right)$, respectively. In each redshift bin, we run {\tt emcee} using 50 walkers for 11,000 steps and discard the first 1,000 steps. 

To prevent low-mass galaxies from biasing the fit due to their greater abundance relative to high-mass galaxies, we apply a mass-dependent weight $W$, which is the inverse of the number density of galaxies at a given mass and redshift. We adopt the stellar mass functions measured by \citet{Davidzon17} to calculate the weights. To account for outliers and mis-classification of star-forming galaxies, we assume an outlier fraction ($P_{out}$) of 5\% and mis-classification probability ($C$) equal to 10\%. 

We compute two variance terms in the likelihood calculation: one term that combines the errors on the effective radii of individual galaxies $e$ and the intrinsic scatter in the size-mass distribution: 
$$\sigma_{\textrm{eff}}^2 = \sigma_{\log \ \textrm{Reff}}^2 + e^2.$$
Another variance term combines the errors on the effective radii of galaxies and the mis-classification probability: 
$$\sigma_{\textrm{mis}}^2 = C^2 + e^2.$$
The likelihood we maximize is
\begin{eqnarray}
\mathcal{L} &= \sum \ln \left[ (1-P_{out}) \cdot W \right] - \frac{1}{2} \ln \left[ 2 \pi \sigma_{\textrm{eff}}^2 \right] \nonumber \\
&- \frac{1}{2} \left[ \log \textrm{R}_{\textrm{eff}} - \left( \log(A) + \alpha \cdot \left[ \log \textrm{M}_{\star} - 10.7\right]\right)\right]^2 / \sigma_{\textrm{eff}}^2 \nonumber \\
&+ \ln P_{out} - \frac{1}{2} \ln \left[ 2 \pi \sigma_{\textrm{mis}}^2 \right] \nonumber \\
&- \frac{1}{2} \left[ \log \textrm{R}_{\textrm{eff}} - \left( \log(A) + \alpha \cdot \left[ \log \textrm{M}_{\star} - 10.7\right]\right)\right]^2 / \sigma_{\textrm{mis}}^2, \nonumber
\end{eqnarray}
where $\log \textrm{M}_{\star}$ here represents the stellar masses of the galaxies in a given redshift bin. We maximize the likelihood separately for the galaxies in each redshift bin and record the best-fitting intercept, slope, and intrinsic scatter. We discuss the results of these fits in the next section. 

\begin{figure*}[t!]
\includegraphics[width=\textwidth]{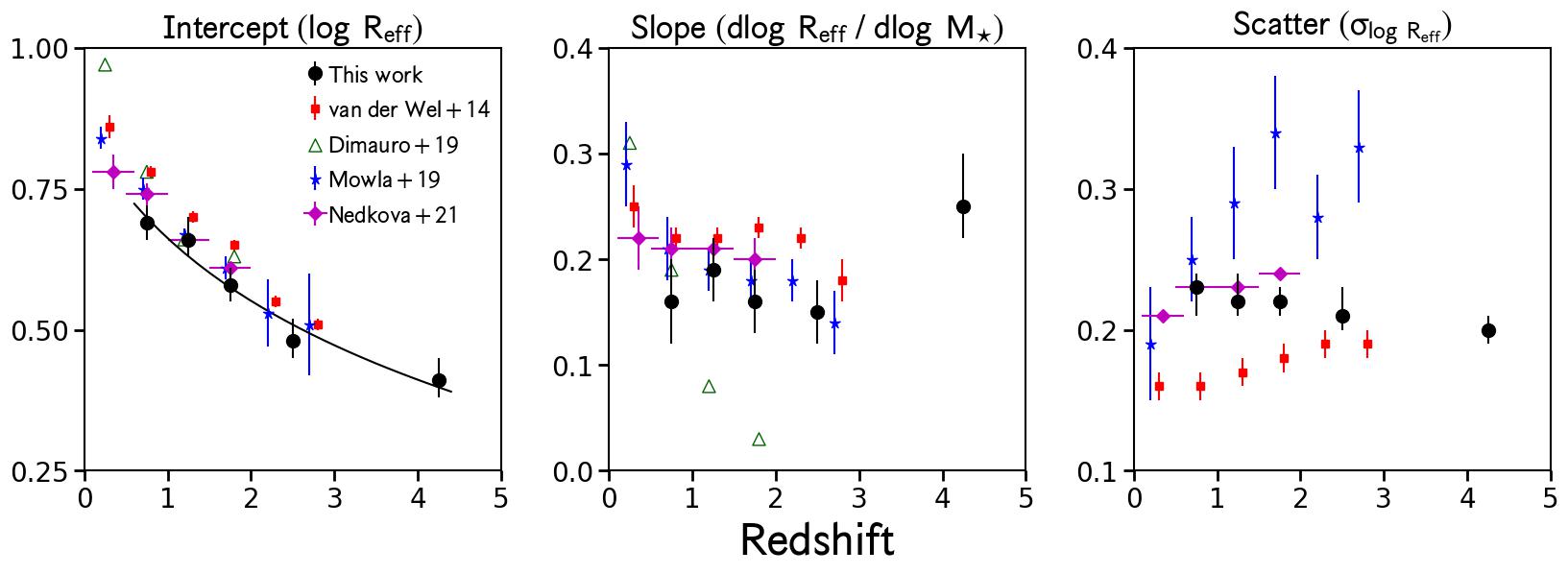}
\caption{Left: best-fitting intercepts at a stellar mass of $5\times 10^{10} \ \textrm{M}_{\odot}$ of the size-mass relations for star-forming galaxies shown in Figure \ref{fig:size_mass} as a function of redshift. Middle: best-fitting slopes of the size-mass relations. Right: Intrinsic scatter in the size-mass relations. Results from the literature are shown in various symbols. Results by \citet{vdw14, Dimauro19, Mowla19} and \citet{Nedkova21} are also shown at the redshifts and stellar masses where these studies and ours overlap. Results from these studies are presented as red squares, empty green triangles, blue stars, and magenta diamonds, respectively. Results obtained by \citet{vdw14} and \citet{Mowla19} are shifted in redshift by +0.05 and -0.05, respectively, for clarity. Please see the text for more details. We find that galaxies become smaller with increasing redshift. The solid line is a power-law fit to the intercept at $\textrm{M}_{\star} = 5\times 10^{10} \ \textrm{M}_{\odot}$ as a function of redshift, $\textrm{R}_{\textrm{eff}} \ / \ \textrm{kpc} = 7.1 (1 + z)^{-0.63}$.The size-mass relation of star-forming galaxies has a steeper slope at $z > 2.5$ than it does at lower redshifts. The scatter in the size-mass relation is relatively constant out to $z=5.5$.  }
\label{fig:size_mass_intercept_slope}
\end{figure*}

% table listing results of size-mass fits
\begin{deluxetable*}{l c c c}[tbh]
\tablecaption{Best-fit Results of the Size-mass Relation of Star-forming Galaxies of the Form $\textrm{R}_{\textrm{eff}} \ / \ \textrm{kpc} = A \left( \textrm{M}_{\star} / 5\times 10^{10} \textrm{M}_{\odot} \right)^{\alpha}$ as Shown in Figures \ref{fig:size_mass} and \ref{fig:size_mass_intercept_slope}\label{tab:size_mass_intercept_slope}}

\tablewidth{\textwidth}

\tablehead{
Redshift range & Intercept $\log(A)$ & Slope ($\alpha$) & $\sigma \left(\log \ \textrm{R}_{\textrm{eff}}\right)$ \\}
\startdata
$0.5 < z < 1.0$ & $0.69_{-0.03}^{+0.03} $ & $0.16_{-0.04}^{+0.04}$ & $0.23_{-0.02}^{+0.01}$ \\
$1.0 < z < 1.5$ & $0.66_{-0.03}^{+0.04}$ & $0.19_{-0.03}^{+0.03}$ & $0.22_{-0.01}^{+0.02}$ \\
$1.5 < z < 2.0$ & $0.58_{-0.03}^{+0.03}$ & $0.16_{-0.03}^{+0.03}$ & $0.22_{-0.01}^{+0.01}$ \\
$2.0 < z < 3.0$ & $0.48_{-0.03}^{+0.04}$ & $0.15_{-0.03}^{+0.03}$ & $0.21_{-0.01}^{+0.02}$ \\
$3.0 < z < 5.5$ & $0.41_{-0.03}^{+0.04}$ & $0.25_{-0.03}^{+0.05}$ & $0.20_{-0.01}^{+0.01}$ \\
\hline
\enddata
\end{deluxetable*}

\section{Results}
\label{sec:results}
\subsection{The Stellar Mass-Size Relation of Star-forming Galaxies out to $z=5.5$}

We show stellar mass against effective radii at rest-frame 5,000\AA\ for both star-forming and quiescent galaxies as blue dots and red squares, respectively, in Figure \ref{fig:size_mass}. The fitted size-mass relations of star-forming galaxies in each redshift bin are shown as solid blue lines. The size-mass relation in the lowest redshift bin is shown as a dashed line in all other redshift bins for comparison. Typical uncertainties in size and mass are shown in the lower right corners of the panels. 

Star-forming galaxies show a clear trend with stellar mass in all redshift bins, such that galaxies increase in size with increasing mass at all redshifts. It is difficult to say how quiescent galaxies behave, given their relatively smaller numbers and limited stellar mass range compared to star-forming galaxies at all redshifts. However, at $z\sim1.25$ and $z\sim1.75$, quiescent galaxies appear to follow a size-mass relation with a steeper slope and lower intercept (by $\sim0.3-0.4$ dex) than those for star-forming galaxies, in qualitative agreement with other studies \citep[e.g.,][]{vdw14, Dimauro19, Mowla19, Nedkova21}. 

The evolution with redshift of the size-mass relation is quantified by applying the method described in Section \ref{sec:fit_size_mass} to the star-forming galaxies in each redshift bin. The best-fitting relations in each redshift bin are shown as blue solid lines in each panel of Figure \ref{fig:size_mass}. The size-mass relation at $z\sim0.7$ is shown as dashed lines in the other redshift bins for comparison. Two results are clear when examining the best-fitting linear relations as a function of redshift. First, the intercept of the size-mass relation decreases with increasing redshift: the solid and dashed lines grow further apart with increasing redshift in Figure \ref{fig:size_mass}. Second, the size-mass relations at $z\sim4.2$ has a steeper slope than at lower redshifts. We examine these results in more detail below. 

The best-fitting parameters and their respective uncertainties are shown as a function of redshift in Figure \ref{fig:size_mass_intercept_slope}. The evolution of the intercept, slope, and scatter with redshift are shown as large black dots with error bars in the left, middle, and right panels, respectively. These best-fit values and their uncertainties are listed in Table \ref{tab:size_mass_intercept_slope}. In the figure, we also show results from other studies that examined the size-mass relation at lower redshifts, namely those by \citet{vdw14, Dimauro19, Mowla19} and \citet{Nedkova21}, in different symbols. We defer a detailed comparison of our results with those obtained by these authors to Section \ref{sec:comparison}. 

The intercept in the size-mass relation evolves with redshift as expected: at a given mass, galaxies become smaller with increasing redshift, as seen in the left panel of Figure \ref{fig:size_mass_intercept_slope}. We quantify the evolution of the intercept with redshift assuming a power law in redshift, consistent with other studies \citep[e.g.,][]{vdw14, Shibuya15, Mowla19}. The best-fitting power-law evolution with redshift at a fixed stellar mass of $5\times10^{10} \ \textrm{M}_{\odot}$ is $\textrm{R}_{\textrm{eff}} \ / \ \textrm{kpc} = (7.1 \pm 0.5) \times (1 + z)^{-0.63 \pm 0.07}$. This result is shown as a black solid line in Figure  \ref{fig:size_mass_intercept_slope} and is consistent to within $2\sigma$ with the results obtained by \citet{vdw14}, \citet{Shibuya15}, and \citet{Mowla19} at similar stellar masses.

The evolution with redshift of the slope is more interesting, as shown in the middle panel of Figure \ref{fig:size_mass_intercept_slope}. At the lowest 4 redshift bins, the slopes range from $0.15 < \alpha < 0.19$ and are roughly constant when taking into account their uncertainties. However, the uncertainties are relatively large, likely due to our small sample size, and the slopes in these lowest four redshift bins are consistent with each other within the uncertainties. At $z\sim4.2$, however, the slope is noticeably higher than those at lower redshifts, though the two slopes are consistent to within $2\sigma$. Overall, the evolution of the slope with redshift is such that the slope appears constant from $\alpha\approx0.15$ to $\alpha\approx0.19$ within our uncertainties at $z\leq 2.5$ and then transitions to a higher slope of $\alpha\approx0.25$ at $z \sim 4.2$. We discuss this evolution with redshift and its potential implications in more detail in Section \ref{sec:implications}. 

Lastly, the evolution with redshift of the intrinsic scatter in the size-mass relation is fairly constant, as seen in the right panel of Figure \ref{fig:size_mass_intercept_slope}. The scatter exhibits small ($\lesssim 0.01$ dex) fluctuations about a typical value of $\sigma_{\log \textrm{Reff}}\approx0.22$ dex. The significance of this result is discussed in greater detail in Section \ref{sec:implications}. 

\subsection{The Morphology-Quenching Relation out to $z\sim4$}
\label{sec:morph_quench}
\begin{figure*}[t!]
\includegraphics[width=\textwidth]{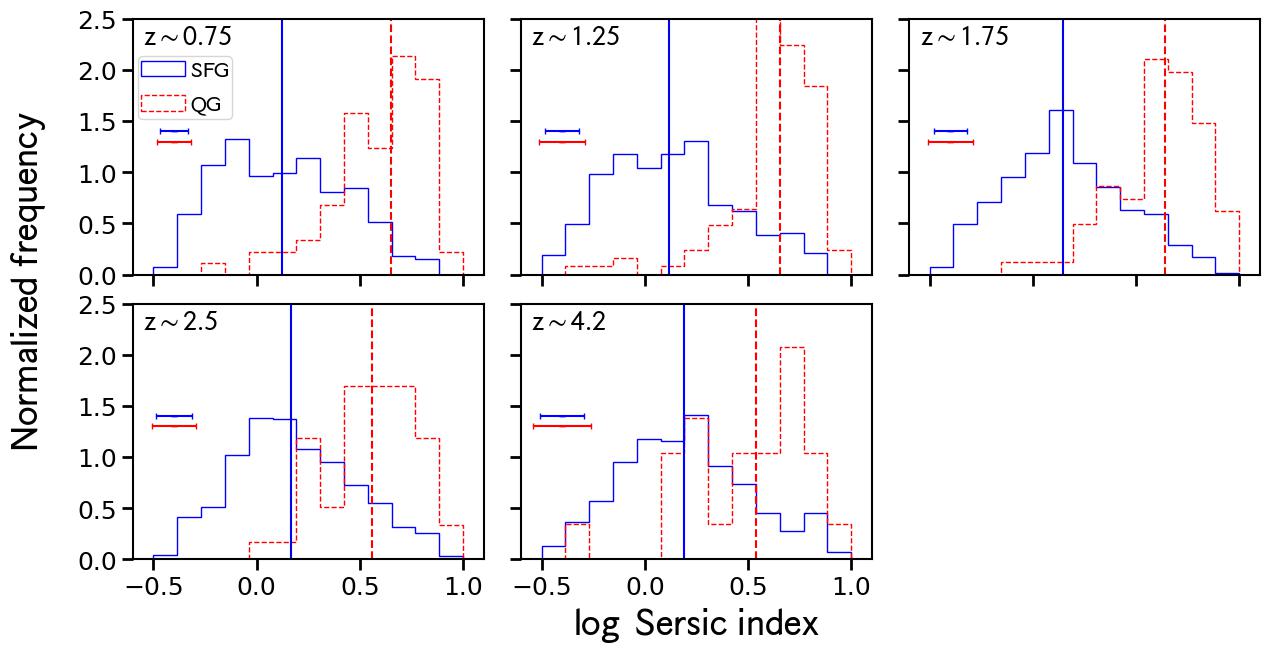}
\caption{S\'{e}rsic index distributions for galaxies in our sample. The distributions of star-forming and quiescent galaxies are shown as blue solid and red dashed histograms, respectively. Median S\'{e}rsic indices of star-forming and quiescent galaxies are indicated by the blue solid and red dashed vertical lines, respectively. Quiescent galaxies have larger S\'{e}rsic indices on average than star-forming galaxies at all redshifts we consider. Despite the small number of quiescent galaxies in our sample at $z\sim4.2$, we find that the S\'{e}rsic index distributions of star-forming and quiescent galaxies are statistically distinguishable (see text). }
\label{fig:sersic_mass}
\end{figure*}

We now examine the distributions of S\'{e}rsic indices for star-forming and quiescent galaxies. We show these distributions as histograms in blue solid and red dashed lines, respectively, in Figure \ref{fig:sersic_mass}. S\'{e}rsic indices in the rest-frame 5,000\AA \ are shown, as is the case for the sizes presented in Figures \ref{fig:size_mass} and \ref{fig:size_mass_intercept_slope}. 

The distributions for both star-forming and quiescent galaxies are clearly separated in all redshift bins. Star-forming galaxies have lower S\'{e}rsic indices on average than quiescent galaxies. The distributions for both kinds of galaxies have a moderate scatter of $\sim0.3$ dex in all redshift bins except for quiescent galaxies in the highest redshfit bin, which have the largest scatter. This increased scatter likely results from very few quiescent galaxies in our sample at the highest redshifts considered. 

The median is taken to be a representative value of the S\'{e}rsic indices of star-forming and quiescent galaxies. The medians of each type of galaxy are shown as blue solid and red dashed vertical lines, respectively, within each panel. Star-forming galaxies have median S\'{e}rsic indices of $n\approx1.3$ at all redshifts, while quiescent galaxies have typical values of $n\sim4.3$ at all except the highest redshifts. In the highest redshift bin, the median S\'{e}rsic index of quiescent galaxies is the lowest, $n\sim3$, compared to what is found at lower redshifts. However, this difference is minor when compared to the distribution at $z\sim2.5$. To determine whether star-forming and quiescent galaxies possess statistically distinguishable S\'{e}rsic index distributions at $z\sim4.2$, we perform a Kolmogorov-Smirnov (KS) test on these distributions. We obtain a KS test statistic of 0.42 and a $p$-value equal to $2.5\times10^{-4}$, which suggests that these distributions are distinct from each other at $>99$\% confidence.

Excluding quiescent galaxies at $z\sim4.2$, our results regarding the typical S\'{e}rsic indices of star-forming and quiescent galaxies are in excellent agreement with what other studies find \citep[cf.][]{Wuyts11, Lang14, Shibuya15}. We note that early results from the CEERS survey also find a morphology-quenching relation out to $z\sim5$ \citep{HuertasCompany23}. These results are derived using machine learning methods, which are completely different from the methods applied here. More recently, rest-frame infrared S\'{e}rsic indices have been measured up to $z = 3$ in the CEERS field by \citet{Martorano23}. These authors find similar median S\'{e}rsic indices for star-forming and quiescent galaxies as we do. 
% AdlV: since we're pointing out that two separate works find results similar to ours, I think the sentence below is superfluous
%It is reassuring that similar qualitative results are obtained using alternative techniques. 

\section{Discussion}
\label{sec:discussion}

\subsection{Comparison with Previous Studies Based on {\it HST} Data}
\label{sec:comparison}

%vdw++14 had to compute color gradients in an averaged sense to find rest frame quantities. We find rest frame quantities galaxy by galaxy, brick by brick, tombstone by tombstone

We compare our results with previous studies in Figure \ref{fig:size_mass_intercept_slope}. These studies also examine the evolution with redshift in the rest-frame optical size-mass relation of star-forming galaxies, using only the {\it HST} data whereas we use both {\it HST} and {\it JWST} data. Overall, we find good agreement among our observations and those of other studies where there is overlap in redshift. Small discrepancies between these studies are likely due to differences in sample selection and size as well as differences in methodology.  

Evolution with redshift of the intercept, slope, and intrinsic scatter of the size-mass relation for star-forming galaxies are shown in the left, middle, and right panels of the figure, respectively. Our results are presented as black dots with error bars. Results from Table 1 by \citet{vdw14} are shown as red squares. Intercepts and slopes from Table 2 by \citet{Dimauro19} are displayed as empty green triangles. These authors did not report uncertainties on their measurements. The results obtained by \citet{Mowla19} and listed in their Tables 1 and 2 are shown as blue stars in the figure. These authors report the observed scatter in the size-mass relation, which is larger than the intrinsic scatter. Lastly, the results from Table 3 by \citet{Nedkova21} are presented as magenta diamonds. These authors report the root mean square error (RMSE) instead of an intrinsic scatter for their size-mass relations. 

Our measurements of the evolution of the size intercept with redshift agree well with earlier studies up to the highest redshifts examined, as seen in the left panel of Figure \ref{fig:size_mass_intercept_slope}. In particular, our results are in good agreement with those by \citet{Mowla19} and \citet{Nedkova21}. Our intercepts are systematically smaller than those measured by \citet{vdw14} and \citet{Dimauro19}, though these offsets are small, $\lesssim$ 0.05 dex. 

At redshifts where these works and ours overlap, our estimates of the slope in the size-mass relation are consistent within errors with those derived by other studies. Our slopes are typically smaller by $\lesssim0.03$ than those measured by other studies. However, the slopes computed by \citet{Dimauro19} show a notable deviation from all other studies. Their slopes decrease with increasing redshift, such that they find $\alpha\sim0.05$ at $z\sim1.6$. In Section 4.2.1 by \citet{Nedkova21}, the authors discuss extensively why the slopes calculated by \citet{Dimauro19} decline steeply with redshift, concluding that differences in sample selection between these two studies are the likely cause. We refer the reader to Appendix B by \citet{Nedkova21} for more details.

%At $z\sim0.7$, our slope lies roughly in-between those measured by other studies, with our uncertainties spanning the range in slopes these authors estimate. From $z\sim1.1$ to $z\sim1.6$, our slopes fall slightly from $\alpha=0.26$ to $\alpha=0.24$, whereas the other studies find slopes more similar to $\alpha\approx0.2$, but with a dispersion that increases with redshift. The slope measured by \citet{vdw14} is most similar to ours at $z\sim1.6$ and that estimated by \citet{Mowla19} is most discrepant, though our determination is consistent with these estimates and that by \citet{Nedkova21} within $2\sigma$.  

At $z\gtrsim3$, we observe an increase in the slope from $\alpha=0.15$ to $\alpha = 0.25$. \citet{vdw14} and \citet{Mowla19} observe mild decreases in their slopes at $2 < z < 3$ and our slope at these redshifts ($\alpha = 0.17$) lies in-between theirs. We discuss the potential significance of discovering a higher slope in the size-mass relation at $z\gtrsim3$ in Section \ref{sec:implications}.

% \citet{vdw14} and \citet{Mowla19} observe a decrease in the slope at $2 < z < 3$, which we recover, albeit with larger uncertainties. At the highest redshifts probed by these studies, our slope ($\alpha = 0.14$) lies in between theirs. At $z\sim4.2$, higher than the redshifts examined by all other studies, we measure the same slope, $\alpha=0.14$, as that observed at $z\sim2.5$. We discuss the potential significance of this discovery in Section \ref{sec:slope}. 

Lastly, the intrinsic scatter in the size-mass relation found here lies between that estimated by other works at overlapping redshifts. Our scatter is nearly constant with redshift, $\sigma \left(\log \ \textrm{R}_{\textrm{eff}}\right) \approx 0.22$. \citet{vdw14} find scatter that is at most $\sim0.1$ dex smaller than ours and \citet{Nedkova21} find RMSE that is consistent with our estimates. As \citet{Mowla19} only report observed, not intrinsic, scatter, their values are considerably larger than what other studies and our own find. However, their observed scatter fluctuates about a typical value of $\sigma \left(\log \ \textrm{R}_{\textrm{eff}}\right) \sim 0.3$, which is roughly constant with redshift and qualitatively similar to what we find.

Offsets between our estimates and those from previous studies may result from a combination of: different redshift bins; different sample sizes (the samples of \citet{vdw14} and \citet{Dimauro19} are approximately 15 and 9 times larger than ours, respectively); different ranges in stellar mass used to measure the size-mass relation; and different methods used to measure rest-frame sizes. Regardless of the causes of these discrepancies, it is reassuring that the same qualitative and similar quantitative evolution in the intercept with redshift is observed when comparing our results with those reported in the literature. 

\subsection{Comparison with Studies Based on {\it JWST} Data}
%\begin{itemize}
%    \item Sun+23 measure size-luminosity relation at $4<z<9$ for star-forming galaxies; find slope that is consistent with that in our highest-z bin
%    \item Ormerod+23 measure size-mass at $0.5 < z < 8.0$, but find different results: they find that the slope decreases with increasing redshift, and their size-redshift results pertain to all galaxies in their sample, complicating any direct comparisons. Sample selection is noticeably different from ours
%\end{itemize}

A handful of studies based on {\it JWST} data have measured the size-mass relation of star-forming galaxies at redshifts that overlap with those investigated in this paper \citep{Ormerod23, Pandya23, Sun23}. However, these studies do not measure all three quantities we use to estimate the size-mass relation (intercept, slope, and scatter) such that direct comparisons between our work and theirs is possible. Therefore, we focus on qualitative comparisons between our results and theirs.

\citet{Sun23} estimate the size-luminosity relation of 347 galaxies at $4 < z < 9.5$ in the CEERS survey. These authors use all seven NIRCam bandpasses in CEERS and perform multi-wavelength morphological fits using {\sc galfitm}, as is done in this work. They infer a slope in the rest-frame optical size-luminosity relation of $0.28\pm0.04$ at $4 <z <9.5$ for the galaxies in their sample with F160W magnitude $<26$; 95\% of the galaxies in our sample are brighter than this magnitude limit. It is encouraging that their size-luminosity slope and our size-mass slope at $3 < z < 5.5$ are consistent within the uncertainties. We can more directly compare our measurements to theirs by computing rest-frame optical luminosities for the subset of our sample at $3 < z < 5.5$. To do so, we take the apparent magnitudes in F277W and F356W for galaxies at $z < 4.2$ and $z > 4.2$, respectively, and convert to absolute magnitudes. We find a slope of $\beta = 0.31 \pm 0.04$ in our size-luminosity relation, which is consistent with the estimates by \citet{Sun23}. At first glance, this comparison would suggest that the slope in the size-mass relation steepens at $z\sim3$ and stays constant up to $z=9.5$, but additional analysis is needed to confirm this.

\citet{Pandya23} investigate the distribution of three-dimensional (3D) shapes of star-forming galaxies at $0.5 < z < 8$ in CEERS. They estimate sizes and axis ratios using both {\sc galfit} and {\sc se++} \citep{Bertin20, Kummel22} and statistically infer the distribution of rest-frame optical 3D morphologies as a function of redshift. Consequently, these authors also infer the mean 3D size-mass relation as a function of redshift. While fits to this relation are not presented, they find that the slope of the size-mass relation decreases with increasing redshift. We find that the slope of the size-mass relation is roughly constant up to $z=3$, then steepens slightly at higher redshift. We cannot conclude what exactly causes this discrepancy, but future work should aim to measure both observed 2D and 3D morphologies and compare.

\citet{Ormerod23} measure the evolution with redshift of galaxy sizes and the size-mass relation in the rest-frame optical for 1,395 galaxies in CEERS. These galaxies are fitted in the NIRCam bandpass that is closest to the rest-frame optical using {\sc galfit}. Galaxies in their sample are selected to have stellar masses greater than $10^{9.5} \ \textrm{M}_{\odot}$, as is done in this study, but they consider galaxies at $0.5 < z \lesssim 8$, whereas our sample is restricted to $0.5 < z < 5.5$. The stellar masses and redshifts used by \citet{Ormerod23} come from different sources than what we use, thus it is difficult to explain why our sample contains $\sim1,000$ more galaxies than theirs despite spanning a smaller range in redshift. With this caveat in mind, we briefly compare our results with theirs. These authors find that galaxy sizes on average decrease with redshift, as we do. However, they consider the evolution with redshift of the average sizes of their sample, instead of that for the intercept in the size-mass relation at a fixed stellar mass, as previous studies and ours have done. Their size-redshift relation (their eq. 5) is consistent with ours to within 2$\sigma$ (see Section \ref{sec:results}). \citet{Ormerod23} find that the sizes of star-forming galaxies increase with stellar mass up to $z\sim6$, though they do not report fitting these relations. Their size-mass relations may be affected by how they separate star-forming and quiescent galaxies. They use the midpoint of the specific star-formation rate distribution in each of their redshift bins to distinguish these populations, whereas we use rest-frame optical colors to do so. A proper comparison of our work with that by \citet{Ormerod23} would necessitate reconstructing their sample, which is beyond the scope of this work.

\subsection{Comparisons with Simulations}
\label{sec: theory and sim}

We now compare our measurements of the size-mass relation as a function of redshift with predictions from simulations. At $z\lesssim3$, where our sample overlaps with prior {\it HST}-based studies, large cosmological hydrodynamical simulations, such as EAGLE \citep{Schaye15, Crain15} and IllustrisTNG \citep{Marinacci18, Naiman18, Nelson18, Pillepich18, Springel18}, are in good agreement with observations \citep{Furlong17, Genel18}. In particular, \citet{Furlong17} report normalizations in their size-mass relations that are within 0.1 dex of those estimated by \citet{vdw14} from $z\sim0$ to $z=2$, and \citet{Genel18} report a slope in the size-mass relation of their star-forming galaxies at $z=2$ that is approximately 0.15. These findings agree well with our results: we find sizes that are also within 0.1 dex of those measured by \citet{vdw14} out to $z=3$, and our slope $\alpha \approx 0.16$ at $z\sim2$ is consistent with what \citet{Genel18} find (see Table \ref{tab:size_mass_intercept_slope}).

At higher redshifts, $z>3$, there is tension between our results and those seen in simulations. \citet{Costantin23} examine how galaxy sizes and the size-mass relation evolve from $z=3$ to $z=6$ in the TNG50 simulation. To do so, these authors produce mock images of their simulated galaxies and include the effects of radiative transfer due to dust and instrumental effects resulting from observing with {\it JWST}. Morphologies are then measured in these mock images and two sets of measurements are presented: one that is intrinsic to the simulation and based on 3D half-mass radii, and one that is based on their mock images. We compare the latter with our results.

\citet{Costantin23} find that the sizes of star-forming galaxies decrease with redshift, as we do, though we find larger sizes at fixed stellar mass. At the stellar mass that corresponds to the intercept of our fits to the size-mass relation, $\log \textrm{M}_{\star} = 10^{10.7} \ \textrm{M}_{\odot}$, we find a typical size of $10^{0.41}$ kpc at $z\sim4.2$, whereas \citet{Costantin23} find a typical size of $10^{0.15}$ kpc at $z=4$. \citet{Costantin23} also find that the slope of the size-mass relation for star-forming galaxies decreases with increasing redshift. At $z=3$ and $z=4$, these authors measure slopes between $a_{\textrm{obs}} = 0.05$ and $a_{\textrm{obs}} = 0.11$ in the F200W and F356W bandpasses. At $z>5$, they find slopes that are negative. At $3 < z < 5.5$, we find that the slope is $0.25^{+0.05}_{-0.03}$, which is in significant disagreement with the TNG50 simulations. Negative slopes in the size-mass relation at $z\gtrsim5$ have been reported in other simulations, such as BLUETIDES \citep{Feng16, Marshall22} and FLARES \citep{Lovell21, Roper22}. \citet{Costantin23} note that the discrepancy between simulated and observed size-mass relations may be attributed to shortcomings in modeling the complex morphologies of high-redshift galaxies when using only one-component S\'{e}rsic models as well as dust obscuration (see also \citealt{Roper22}).

\subsection{Implications Regarding the Evolution of Star-forming Galaxies}
\label{sec:implications}

Models of disk galaxy formation and evolution often assume proportional relationships between the properties of dark matter halos and those of the galaxies they host \citep[e.g.,][]{Fall80, MMW98}. These proportional relations arise as gas in galaxies is expected to inherit the specific angular momentum from the host dark matter halo. We briefly discuss how our results fit into this paradigm below. 

The intrinsic scatter in our measured size-mass relation is nearly constant at $\sigma \left(\log \ \textrm{R}_{\textrm{eff}}\right) \approx 0.22$ from $z\sim0.7$ to $z\sim4.2$. This value is strikingly similar to the predicted scatter of 0.25 dex (in log base 10) in the halo spin parameter \citep[e.g.,][]{Bullock01, Maccio08}. This finding suggests that the sizes of galaxies are set by the sizes of their dark matter halos since $z=5.5$. We consider the connection between galaxy and dark matter halo properties further. 

We find that the slope in the size-mass relation is approximately constant at $\alpha\approx0.16$ up to $z\sim3$ and increases slightly to $\alpha=0.25$ at higher redshift. \citet{Shen03} argue that if the ratio of galaxy mass to halo mass were constant, the observed slope would be steeper, $\alpha = 1/3$. This argument assumes that galaxy sizes are proportional to halo sizes. At $0 < z <  3$, disk galaxies are indeed observed to have sizes that are proportional to those of their halos \citep{Kravtsov13, Huang17}. Assuming this relation at $z\sim4.2$, we use Equation 23 by \citet{Shen03} to derive the stellar mass-halo mass relation, finding $\textrm{M}_{\star} / \textrm{M}_{h} \propto M_{h}^{1/(3\alpha) - 1}$. From our results, this suggests that $\textrm{M}_{\star} / \textrm{M}_{h} \propto M_{h}^{\gamma \sim 1}$ at $z\lesssim3$ and $\textrm{M}_{\star} / \textrm{M}_{h} \propto M_{h}^{\gamma \sim 0.3}$ at $z\gtrsim3$. Our inferred slopes in the stellar mass - halo mass relation are consistent with measurements thereof at $z<3$ for lower halo masses \citep{Behroozi13, Moster13, Girelli20}. 

%At first glance, these results suggest that, at stellar masses $\textrm{M}_{\star} \geq 10^{9.5} \ \textrm{M}_{\odot}$, the stellar mass - halo mass relation is shallow at low redshifts and considerably steeper at high redshifts. However, the stellar mass - halo mass relation (SHMR) is often parameterized by a double power law such that there are separate slopes at low and high halo masses. At $0 < z < 4$, galaxies of stellar mass $10 < \log \left( \textrm{M}_{\star} \ / \ \textrm{M}_{\odot}\right) < 11$ have halo masses of $\textrm{M}_h \sim 10^{12-13} \textrm{M}_{\odot}$ \citep{Behroozi13, Moster13}. The halo mass at which the double power law transitions evolves with redshift, such that it increases from $\textrm{M}_h \sim 10^{12} \ \textrm{M}_{\odot}$ at $z=0$ to $\textrm{M}_h \sim 10^{12.5} \ \textrm{M}_{\odot}$ at $z=4$ \citep{Moster13, Girelli20}. Therefore, our galaxy sample lies in different regimes of the SHMR at the extreme ends of our redshift distribution: at the lowest redshifts, our sample primarily lies in the high-mass portion of the SHMR, while at the highest redshifts, our sample lies mainly in the low-mass portion of the SHMR. Now taking these regimes into account, our predictions of the slopes of the SHMR at different redshifts are consistent with recent observations at high redshift \citep[see, e.g., Table 2 in][]{Girelli20}. 

The results we have discussed thus far imply that star-forming galaxies on average at $0.5 < z < 5.5$ inherit the specific angular momenta of their host DM halos. This suggests that the stellar kinematics of these galaxies may resemble what is expected of rotating disks. However, a number of studies find that high-redshift galaxies are likely not rotating, oblate disks, but rather prolate, cigar-shaped systems \citep{vdw14b, Zhang19, Pandya23}. These findings are also supported by simulations \citep{Tomassetti16, VegaFerrero23}. While the results from these studies may complicate the interpretation of our results above, the possibility that high-redshift galaxies are predominantly prolate may explain why we observe a steeper size-mass slope at $z>3$ than at lower redshifts. 

A steeper slope in the size-mass relation with respect to redshift may arise due to either A) low-mass galaxies decreasing in size with redshift faster than high-mass galaxies, or B) high-mass galaxies increasing in size with redshift faster than low-mass galaxies. As can be seen in Figure \ref{fig:size_mass_intercept_slope}, sizes at $\textrm{M}_{\star} = 5\times10^{10} \ \textrm{M}_{\odot}$ decrease with redshift, hence the latter explanation above is unlikely. \citet{Pandya23} find that low-mass galaxies are most likely prolate at high redshift, with prolate fractions of $\sim80\%$ at $3 <z <8$. The transition from oblate disks at low redshift to prolate systems at high redshift may be accompanied by transitions in the sizes of such galaxies from large to small.

\section{Conclusions}
\label{sec:conclusions}

We use the early release {\it JWST} data taken by the CEERS survey, in combination with available {\it HST} data, to examine the size-mass relation of star-forming galaxies and compare the morphologies of star-forming and quiescent galaxies in the rest-frame optical at $0.5 < z < 5.5$. We fit the sizes of 2,450 galaxies across 13 bandpasses and derive sizes and S\'{e}rsic indices in the rest-frame 5,000\AA \ using a simultaneous multi-wavelength fitting approach. 

The main conclusions we find from our investigation of the mass-size relation are summarized as follows:

\begin{enumerate}
    \item The slope of the size-mass relation ($\textrm{d}\log \textrm{R}_{\textrm{eff}} \ / \ \textrm{d}\log \textrm{M}_{\star}$) appears to be constant out to $z\lesssim3$ and steepens slightly at $z\gtrsim3$ (middle panel in Figure \ref{fig:size_mass_intercept_slope}). Our measurements are consistent with previous work out to $z\sim3$ as well as some studies at higher redshifts than what is considered here. 

    \item The intercept of the size-mass relation ($\log(A)$) continues to decline at a rate similar to what other studies find out to $z=5.5$ (left panel of Figure \ref{fig:size_mass_intercept_slope}). Our measurements agree well with those in the literature at lower redshifts and are consistent despite our smaller sample size by a factor of at least 9.

    \item The scatter of the size-mass relation ($\sigma \left(\log \ \textrm{R}_{\textrm{eff}}\right)$) is roughly constant out to $z=5.5$ (right panel in Figure \ref{fig:size_mass_intercept_slope}). This suggests a tight connection between the properties of star-forming galaxies and their dark matter halos, which may suggest the prevalence of disk galaxies at high redshift. However, our observations of the evolution of the slope with redshift may also be consistent with the occurrence of prolate galaxies at similar redshifts. 
    
    \item Star-forming and quiescent galaxies have clearly separate distributions in S\'{e}rsic index. Star-forming and quiescent galaxies have typical S\'{e}rsic indices $n \sim 1.3$ and $n \sim 4.3$, respectively, at all redshifts we consider. 
\end{enumerate}

This pilot study of galaxy morphologies at high redshift, with the unparalleled combination of {\it{JWST}} and {\it{HST}} data, extends the characterization of the size-mass relation of star-forming galaxies and the morphology-quenching relation at high stellar mass out to $z = 5.5$. Future work would benefit from combining data from CEERS with other large {\it JWST} surveys, such as COSMOS-Web \citep{Casey22} and UNCOVER \citep{Bezanson22}.

%% Also note that the akcnowlodgment environment does not support long amounts of text. If you have a lot of people and institutions to acknowledge, do not use this command. Instead, create a new \section{Acknowledgments}.

\section{Acknowledgements}
\begin{acknowledgments}
EW and AdlV thank Amanda Pagul, Javier Sanchez, and Ian McConachie for helpful conversations and advice throughout the writing of this paper. EW would also like to thank Aimee Ward and Brian Crowell for their continued support and encouragement throughout the preparation of this work.
\end{acknowledgments}


\vspace{5mm}
\facilities{{\it HST}, {\it JWST}}


\software{astropy \citep{2013A&A...558A..33A,2018AJ....156..123A}, LMFIT \citep{Newville14}, matplotlib \citep{Hunter07}, numpy \citep{harris20}, scipy \citep{Scipy20}, photutils \citep{larry_bradley_2022_6825092}, emcee \citep{ForemanMackey13}, {\sc galfitm} \citep{Barden12, Haussler13, Vika13, Haussler22}}




\bibliography{sample631}{}

\begin{thebibliography}{}
\expandafter\ifx\csname natexlab\endcsname\relax\def\natexlab#1{#1}\fi
\providecommand{\url}[1]{\href{#1}{#1}}
\providecommand{\dodoi}[1]{doi:~\href{http://doi.org/#1}{\nolinkurl{#1}}}
\providecommand{\doeprint}[1]{\href{http://ascl.net/#1}{\nolinkurl{http://ascl.net/#1}}}
\providecommand{\doarXiv}[1]{\href{https://arxiv.org/abs/#1}{\nolinkurl{https://arxiv.org/abs/#1}}}

\bibitem[{{Abramowitz} \& {Stegun}(1965)}]{AbramowitzStegun65}
{Abramowitz}, M., \& {Stegun}, I.~A. 1965, {Handbook of mathematical functions
  with formulas, graphs, and mathematical tables}

\bibitem[{{Astropy Collaboration} {et~al.}(2013){Astropy Collaboration},
  {Robitaille}, {Tollerud}, {Greenfield}, {Droettboom}, {Bray}, {Aldcroft},
  {Davis}, {Ginsburg}, {Price-Whelan}, {Kerzendorf}, {Conley}, {Crighton},
  {Barbary}, {Muna}, {Ferguson}, {Grollier}, {Parikh}, {Nair}, {Unther},
  {Deil}, {Woillez}, {Conseil}, {Kramer}, {Turner}, {Singer}, {Fox}, {Weaver},
  {Zabalza}, {Edwards}, {Azalee Bostroem}, {Burke}, {Casey}, {Crawford},
  {Dencheva}, {Ely}, {Jenness}, {Labrie}, {Lim}, {Pierfederici}, {Pontzen},
  {Ptak}, {Refsdal}, {Servillat}, \& {Streicher}}]{2013A&A...558A..33A}
{Astropy Collaboration}, {Robitaille}, T.~P., {Tollerud}, E.~J., {et~al.} 2013,
  \aap, 558, A33, \dodoi{10.1051/0004-6361/201322068}

\bibitem[{{Astropy Collaboration} {et~al.}(2018){Astropy Collaboration},
  {Price-Whelan}, {Sip{\H{o}}cz}, {G{\"u}nther}, {Lim}, {Crawford}, {Conseil},
  {Shupe}, {Craig}, {Dencheva}, {Ginsburg}, {VanderPlas}, {Bradley},
  {P{\'e}rez-Su{\'a}rez}, {de Val-Borro}, {Aldcroft}, {Cruz}, {Robitaille},
  {Tollerud}, {Ardelean}, {Babej}, {Bach}, {Bachetti}, {Bakanov}, {Bamford},
  {Barentsen}, {Barmby}, {Baumbach}, {Berry}, {Biscani}, {Boquien}, {Bostroem},
  {Bouma}, {Brammer}, {Bray}, {Breytenbach}, {Buddelmeijer}, {Burke},
  {Calderone}, {Cano Rodr{\'\i}guez}, {Cara}, {Cardoso}, {Cheedella}, {Copin},
  {Corrales}, {Crichton}, {D'Avella}, {Deil}, {Depagne}, {Dietrich}, {Donath},
  {Droettboom}, {Earl}, {Erben}, {Fabbro}, {Ferreira}, {Finethy}, {Fox},
  {Garrison}, {Gibbons}, {Goldstein}, {Gommers}, {Greco}, {Greenfield},
  {Groener}, {Grollier}, {Hagen}, {Hirst}, {Homeier}, {Horton}, {Hosseinzadeh},
  {Hu}, {Hunkeler}, {Ivezi{\'c}}, {Jain}, {Jenness}, {Kanarek}, {Kendrew},
  {Kern}, {Kerzendorf}, {Khvalko}, {King}, {Kirkby}, {Kulkarni}, {Kumar},
  {Lee}, {Lenz}, {Littlefair}, {Ma}, {Macleod}, {Mastropietro}, {McCully},
  {Montagnac}, {Morris}, {Mueller}, {Mumford}, {Muna}, {Murphy}, {Nelson},
  {Nguyen}, {Ninan}, {N{\"o}the}, {Ogaz}, {Oh}, {Parejko}, {Parley}, {Pascual},
  {Patil}, {Patil}, {Plunkett}, {Prochaska}, {Rastogi}, {Reddy Janga},
  {Sabater}, {Sakurikar}, {Seifert}, {Sherbert}, {Sherwood-Taylor}, {Shih},
  {Sick}, {Silbiger}, {Singanamalla}, {Singer}, {Sladen}, {Sooley},
  {Sornarajah}, {Streicher}, {Teuben}, {Thomas}, {Tremblay}, {Turner},
  {Terr{\'o}n}, {van Kerkwijk}, {de la Vega}, {Watkins}, {Weaver}, {Whitmore},
  {Woillez}, {Zabalza}, \& {Astropy Contributors}}]{2018AJ....156..123A}
{Astropy Collaboration}, {Price-Whelan}, A.~M., {Sip{\H{o}}cz}, B.~M., {et~al.}
  2018, \aj, 156, 123, \dodoi{10.3847/1538-3881/aabc4f}

\bibitem[{{Bagley} {et~al.}(2023){Bagley}, {Finkelstein}, {Koekemoer},
  {Ferguson}, {Arrabal Haro}, {Dickinson}, {Kartaltepe}, {Papovich},
  {P{\'e}rez-Gonz{\'a}lez}, {Pirzkal}, {Somerville}, {Willmer}, {Yang}, {Yung},
  {Fontana}, {Grazian}, {Grogin}, {Hirschmann}, {Kewley}, {Kirkpatrick},
  {Kocevski}, {Lotz}, {Medrano}, {Morales}, {Pentericci}, {Ravindranath},
  {Trump}, {Wilkins}, {Calabr{\`o}}, {Cooper}, {Costantin}, {de la Vega},
  {Hilbert}, {Hutchison}, {Larson}, {Lucas}, {McGrath}, {Ryan}, {Wang}, \&
  {Wuyts}}]{Bagley23}
{Bagley}, M.~B., {Finkelstein}, S.~L., {Koekemoer}, A.~M., {et~al.} 2023,
  \apjl, 946, L12, \dodoi{10.3847/2041-8213/acbb08}

\bibitem[{{Baldry} {et~al.}(2006){Baldry}, {Balogh}, {Bower}, {Glazebrook},
  {Nichol}, {Bamford}, \& {Budavari}}]{Baldry06}
{Baldry}, I.~K., {Balogh}, M.~L., {Bower}, R.~G., {et~al.} 2006, \mnras, 373,
  469, \dodoi{10.1111/j.1365-2966.2006.11081.x}

\bibitem[{{Barden} {et~al.}(2012){Barden}, {H{\"a}u{\ss}ler}, {Peng},
  {McIntosh}, \& {Guo}}]{Barden12}
{Barden}, M., {H{\"a}u{\ss}ler}, B., {Peng}, C.~Y., {McIntosh}, D.~H., \&
  {Guo}, Y. 2012, \mnras, 422, 449, \dodoi{10.1111/j.1365-2966.2012.20619.x}

\bibitem[{{Barro} {et~al.}(2023){Barro}, {Perez-Gonzalez}, {Kocevski},
  {McGrath}, {Trump}, {Simons}, {Somerville}, {Yung}, {Arrabal Haro}, {Bagley},
  {Cleri}, {Costantin}, {Davis}, {Dickinson}, {Finkelstein}, {Giavalisco},
  {Gomez-Guijarro}, {Hathi}, {Hirschmann}, {Akins}, {Holwerda},
  {Huertas-Company}, {Lucas}, {Papovich}, {Seille}, {Tacchella}, {Wilkins}, {de
  la Vega}, {Yang}, \& {Zavala}}]{Barro23}
{Barro}, G., {Perez-Gonzalez}, P.~G., {Kocevski}, D.~D., {et~al.} 2023, arXiv
  e-prints, arXiv:2305.14418, \dodoi{10.48550/arXiv.2305.14418}

\bibitem[{{Behroozi} {et~al.}(2013){Behroozi}, {Wechsler}, \&
  {Conroy}}]{Behroozi13}
{Behroozi}, P.~S., {Wechsler}, R.~H., \& {Conroy}, C. 2013, \apj, 770, 57,
  \dodoi{10.1088/0004-637X/770/1/57}

\bibitem[{{Bertin} {et~al.}(2020){Bertin}, {Schefer}, {Apostolakos},
  {{\'A}lvarez-Ayll{\'o}n}, {Dubath}, \& {K{\"u}mmel}}]{Bertin20}
{Bertin}, E., {Schefer}, M., {Apostolakos}, N., {et~al.} 2020, in Astronomical
  Society of the Pacific Conference Series, Vol. 527, Astronomical Data
  Analysis Software and Systems XXIX, ed. R.~{Pizzo}, E.~R. {Deul}, J.~D.
  {Mol}, J.~{de Plaa}, \& H.~{Verkouter}, 461

\bibitem[{{Bezanson} {et~al.}(2009){Bezanson}, {van Dokkum}, {Tal},
  {Marchesini}, {Kriek}, {Franx}, \& {Coppi}}]{Bezanson09}
{Bezanson}, R., {van Dokkum}, P.~G., {Tal}, T., {et~al.} 2009, \apj, 697, 1290,
  \dodoi{10.1088/0004-637X/697/2/1290}

\bibitem[{{Bezanson} {et~al.}(2022){Bezanson}, {Labbe}, {Whitaker}, {Leja},
  {Price}, {Franx}, {Brammer}, {Marchesini}, {Zitrin}, {Wang}, {Weaver},
  {Furtak}, {Atek}, {Coe}, {Cutler}, {Dayal}, {van Dokkum}, {Feldmann},
  {Forster Schreiber}, {Fujimoto}, {Geha}, {Glazebrook}, {de Graaff}, {Greene},
  {Juneau}, {Kassin}, {Kriek}, {Khullar}, {Maseda}, {Mowla}, {Muzzin},
  {Nanayakkara}, {Nelson}, {Oesch}, {Pacifici}, {Pan}, {Papovich}, {Setton},
  {Shapley}, {Smit}, {Stefanon}, {Taylor}, \& {Williams}}]{Bezanson22}
{Bezanson}, R., {Labbe}, I., {Whitaker}, K.~E., {et~al.} 2022, arXiv e-prints,
  arXiv:2212.04026, \dodoi{10.48550/arXiv.2212.04026}

\bibitem[{Bradley {et~al.}(2022)Bradley, Sipőcz, Robitaille, Tollerud,
  Vinícius, Deil, Barbary, Wilson, Busko, Donath, Günther, Cara, Lim,
  Meßlinger, Conseil, Bostroem, Droettboom, Bray, Bratholm, Barentsen, Craig,
  Rathi, Pascual, Perren, Georgiev, de~Val-Borro, Kerzendorf, Bach, Quint, \&
  Souchereau}]{larry_bradley_2022_6825092}
Bradley, L., Sipőcz, B., Robitaille, T., {et~al.} 2022, astropy/photutils:
  1.5.0, 1.5.0,  Zenodo, \dodoi{10.5281/zenodo.6825092}

\bibitem[{{Brammer} {et~al.}(2008){Brammer}, {van Dokkum}, \&
  {Coppi}}]{Brammer08}
{Brammer}, G.~B., {van Dokkum}, P.~G., \& {Coppi}, P. 2008, \apj, 686, 1503,
  \dodoi{10.1086/591786}

\bibitem[{{Bruce} {et~al.}(2012){Bruce}, {Dunlop}, {Cirasuolo}, {McLure},
  {Targett}, {Bell}, {Croton}, {Dekel}, {Faber}, {Ferguson}, {Grogin},
  {Kocevski}, {Koekemoer}, {Koo}, {Lai}, {Lotz}, {McGrath}, {Newman}, \& {van
  der Wel}}]{Bruce12}
{Bruce}, V.~A., {Dunlop}, J.~S., {Cirasuolo}, M., {et~al.} 2012, \mnras, 427,
  1666, \dodoi{10.1111/j.1365-2966.2012.22087.x}

\bibitem[{{Bruzual} \& {Charlot}(2003)}]{BC03}
{Bruzual}, G., \& {Charlot}, S. 2003, \mnras, 344, 1000,
  \dodoi{10.1046/j.1365-8711.2003.06897.x}

\bibitem[{{Buitrago} {et~al.}(2008){Buitrago}, {Trujillo}, {Conselice},
  {Bouwens}, {Dickinson}, \& {Yan}}]{Buitrago08}
{Buitrago}, F., {Trujillo}, I., {Conselice}, C.~J., {et~al.} 2008, \apjl, 687,
  L61, \dodoi{10.1086/592836}

\bibitem[{{Bullock} {et~al.}(2001){Bullock}, {Dekel}, {Kolatt}, {Kravtsov},
  {Klypin}, {Porciani}, \& {Primack}}]{Bullock01}
{Bullock}, J.~S., {Dekel}, A., {Kolatt}, T.~S., {et~al.} 2001, \apj, 555, 240,
  \dodoi{10.1086/321477}

\bibitem[{{Calzetti} {et~al.}(2000){Calzetti}, {Armus}, {Bohlin}, {Kinney},
  {Koornneef}, \& {Storchi-Bergmann}}]{Calzetti00}
{Calzetti}, D., {Armus}, L., {Bohlin}, R.~C., {et~al.} 2000, \apj, 533, 682,
  \dodoi{10.1086/308692}

\bibitem[{{Carollo} {et~al.}(2013){Carollo}, {Bschorr}, {Renzini}, {Lilly},
  {Capak}, {Cibinel}, {Ilbert}, {Onodera}, {Scoville}, {Cameron}, {Mobasher},
  {Sanders}, \& {Taniguchi}}]{Carollo13}
{Carollo}, C.~M., {Bschorr}, T.~J., {Renzini}, A., {et~al.} 2013, \apj, 773,
  112, \dodoi{10.1088/0004-637X/773/2/112}

\bibitem[{{Casey} {et~al.}(2022){Casey}, {Kartaltepe}, {Drakos}, {Franco},
  {Harish}, {Paquereau}, {Ilbert}, {Rose}, {Cox}, {Nightingale}, {Robertson},
  {Silverman}, {Koekemoer}, {Massey}, {McCracken}, {Rhodes}, {Akins},
  {Amvrosiadis}, {Arango-Toro}, {Bagley}, {Bongiorno}, {Capak}, {Champagne},
  {Chartab}, {Chavez Ortiz}, {Chworowsky}, {Cooke}, {Cooper}, {Darvish},
  {Ding}, {Faisst}, {Finkelstein}, {Fujimoto}, {Gentile}, {Gillman}, {Gould},
  {Gozaliasl}, {Hayward}, {He}, {Hemmati}, {Hirschmann}, {Jahnke}, {Jin},
  {Khostovan}, {Kokorev}, {Lambrides}, {Laigle}, {Larson}, {Leung}, {Liu},
  {Liaudat}, {Long}, {Magdis}, {Mahler}, {Mainieri}, {Manning}, {Maraston},
  {Martin}, {McCleary}, {McKinney}, {McPartland}, {Mobasher}, {Pattnaik},
  {Renzini}, {Rich}, {Sanders}, {Sattari}, {Scognamiglio}, {Scoville}, {Sheth},
  {Shuntov}, {Sparre}, {Suzuki}, {Talia}, {Toft}, {Trakhtenbrot}, {Urry},
  {Valentino}, {Vanderhoof}, {Vardoulaki}, {Weaver}, {Whitaker}, {Wilkins},
  {Yang}, \& {Zavala}}]{Casey22}
{Casey}, C.~M., {Kartaltepe}, J.~S., {Drakos}, N.~E., {et~al.} 2022, arXiv
  e-prints, arXiv:2211.07865, \dodoi{10.48550/arXiv.2211.07865}

\bibitem[{{Chabrier}(2003)}]{Chabrier03}
{Chabrier}, G. 2003, \pasp, 115, 763, \dodoi{10.1086/376392}

\bibitem[{{Conroy} \& {Gunn}(2010)}]{Conroy10}
{Conroy}, C., \& {Gunn}, J.~E. 2010, \apj, 712, 833,
  \dodoi{10.1088/0004-637X/712/2/833}

\bibitem[{{Conselice} {et~al.}(2011){Conselice}, {Bluck}, {Buitrago}, {Bauer},
  {Gr{\"u}tzbauch}, {Bouwens}, {Bevan}, {Mortlock}, {Dickinson}, {Daddi},
  {Yan}, {Scott}, {Chapman}, {Chary}, {Ferguson}, {Giavalisco}, {Grogin},
  {Illingworth}, {Jogee}, {Koekemoer}, {Lucas}, {Mobasher}, {Moustakas},
  {Papovich}, {Ravindranath}, {Siana}, {Teplitz}, {Trujillo}, {Urry}, \&
  {Weinzirl}}]{Conselice11}
{Conselice}, C.~J., {Bluck}, A.~F.~L., {Buitrago}, F., {et~al.} 2011, \mnras,
  413, 80, \dodoi{10.1111/j.1365-2966.2010.18113.x}

\bibitem[{{Costantin} {et~al.}(2023){Costantin}, {P{\'e}rez-Gonz{\'a}lez},
  {Vega-Ferrero}, {Huertas-Company}, {Bisigello}, {Buitrago}, {Bagley},
  {Cleri}, {Cooper}, {Finkelstein}, {Holwerda}, {Kartaltepe}, {Koekemoer},
  {Nelson}, {Papovich}, {Pillepich}, {Pirzkal}, {Tacchella}, \&
  {Yung}}]{Costantin23}
{Costantin}, L., {P{\'e}rez-Gonz{\'a}lez}, P.~G., {Vega-Ferrero}, J., {et~al.}
  2023, \apj, 946, 71, \dodoi{10.3847/1538-4357/acb926}

\bibitem[{{Crain} {et~al.}(2015){Crain}, {Schaye}, {Bower}, {Furlong},
  {Schaller}, {Theuns}, {Dalla Vecchia}, {Frenk}, {McCarthy}, {Helly},
  {Jenkins}, {Rosas-Guevara}, {White}, \& {Trayford}}]{Crain15}
{Crain}, R.~A., {Schaye}, J., {Bower}, R.~G., {et~al.} 2015, \mnras, 450, 1937,
  \dodoi{10.1093/mnras/stv725}

\bibitem[{{Dahlen} {et~al.}(2013){Dahlen}, {Mobasher}, {Faber}, {Ferguson},
  {Barro}, {Finkelstein}, {Finlator}, {Fontana}, {Gruetzbauch}, {Johnson},
  {Pforr}, {Salvato}, {Wiklind}, {Wuyts}, {Acquaviva}, {Dickinson}, {Guo},
  {Huang}, {Huang}, {Newman}, {Bell}, {Conselice}, {Galametz}, {Gawiser},
  {Giavalisco}, {Grogin}, {Hathi}, {Kocevski}, {Koekemoer}, {Koo}, {Lee},
  {McGrath}, {Papovich}, {Peth}, {Ryan}, {Somerville}, {Weiner}, \&
  {Wilson}}]{Dahlen13}
{Dahlen}, T., {Mobasher}, B., {Faber}, S.~M., {et~al.} 2013, \apj, 775, 93,
  \dodoi{10.1088/0004-637X/775/2/93}

\bibitem[{{Davidzon} {et~al.}(2017){Davidzon}, {Ilbert}, {Laigle}, {Coupon},
  {McCracken}, {Delvecchio}, {Masters}, {Capak}, {Hsieh}, {Le F{\`e}vre},
  {Tresse}, {Bethermin}, {Chang}, {Faisst}, {Le Floc'h}, {Steinhardt}, {Toft},
  {Aussel}, {Dubois}, {Hasinger}, {Salvato}, {Sanders}, {Scoville}, \&
  {Silverman}}]{Davidzon17}
{Davidzon}, I., {Ilbert}, O., {Laigle}, C., {et~al.} 2017, \aap, 605, A70,
  \dodoi{10.1051/0004-6361/201730419}

\bibitem[{{Davis} {et~al.}(2007){Davis}, {Guhathakurta}, {Konidaris}, {Newman},
  {Ashby}, {Biggs}, {Barmby}, {Bundy}, {Chapman}, {Coil}, {Conselice},
  {Cooper}, {Croton}, {Eisenhardt}, {Ellis}, {Faber}, {Fang}, {Fazio},
  {Georgakakis}, {Gerke}, {Goss}, {Gwyn}, {Harker}, {Hopkins}, {Huang},
  {Ivison}, {Kassin}, {Kirby}, {Koekemoer}, {Koo}, {Laird}, {Le Floc'h}, {Lin},
  {Lotz}, {Marshall}, {Martin}, {Metevier}, {Moustakas}, {Nandra}, {Noeske},
  {Papovich}, {Phillips}, {Rich}, {Rieke}, {Rigopoulou}, {Salim},
  {Schiminovich}, {Simard}, {Smail}, {Small}, {Weiner}, {Willmer}, {Willner},
  {Wilson}, {Wright}, \& {Yan}}]{Davis07}
{Davis}, M., {Guhathakurta}, P., {Konidaris}, N.~P., {et~al.} 2007, \apjl, 660,
  L1, \dodoi{10.1086/517931}

\bibitem[{{Dimauro} {et~al.}(2019){Dimauro}, {Huertas-Company}, {Daddi},
  {P{\'e}rez-Gonz{\'a}lez}, {Bernardi}, {Caro}, {Cattaneo}, {H{\"a}u{\ss}ler},
  {Kuchner}, {Shankar}, {Barro}, {Buitrago}, {Faber}, {Kocevski}, {Koekemoer},
  {Koo}, {Mei}, {Peletier}, {Primack}, {Rodriguez-Puebla}, {Salvato}, \&
  {Tuccillo}}]{Dimauro19}
{Dimauro}, P., {Huertas-Company}, M., {Daddi}, E., {et~al.} 2019, \mnras, 489,
  4135, \dodoi{10.1093/mnras/stz2421}

\bibitem[{{Dutton} {et~al.}(2007){Dutton}, {van den Bosch}, {Dekel}, \&
  {Courteau}}]{Dutton07}
{Dutton}, A.~A., {van den Bosch}, F.~C., {Dekel}, A., \& {Courteau}, S. 2007,
  \apj, 654, 27, \dodoi{10.1086/509314}

\bibitem[{{El-Badry} {et~al.}(2016){El-Badry}, {Wetzel}, {Geha}, {Hopkins},
  {Kere{\v{s}}}, {Chan}, \& {Faucher-Gigu{\`e}re}}]{ElBadry16}
{El-Badry}, K., {Wetzel}, A., {Geha}, M., {et~al.} 2016, \apj, 820, 131,
  \dodoi{10.3847/0004-637X/820/2/131}

\bibitem[{{Faisst} {et~al.}(2017){Faisst}, {Carollo}, {Capak}, {Tacchella},
  {Renzini}, {Ilbert}, {McCracken}, \& {Scoville}}]{Faisst17}
{Faisst}, A.~L., {Carollo}, C.~M., {Capak}, P.~L., {et~al.} 2017, \apj, 839,
  71, \dodoi{10.3847/1538-4357/aa697a}

\bibitem[{{Fall} \& {Efstathiou}(1980)}]{Fall80}
{Fall}, S.~M., \& {Efstathiou}, G. 1980, \mnras, 193, 189,
  \dodoi{10.1093/mnras/193.2.189}

\bibitem[{{Feng} {et~al.}(2016){Feng}, {Di-Matteo}, {Croft}, {Bird},
  {Battaglia}, \& {Wilkins}}]{Feng16}
{Feng}, Y., {Di-Matteo}, T., {Croft}, R.~A., {et~al.} 2016, \mnras, 455, 2778,
  \dodoi{10.1093/mnras/stv2484}

\bibitem[{{Ferguson} {et~al.}(2004){Ferguson}, {Dickinson}, {Giavalisco},
  {Kretchmer}, {Ravindranath}, {Idzi}, {Taylor}, {Conselice}, {Fall},
  {Gardner}, {Livio}, {Madau}, {Moustakas}, {Papovich}, {Somerville},
  {Spinrad}, \& {Stern}}]{Ferguson04}
{Ferguson}, H.~C., {Dickinson}, M., {Giavalisco}, M., {et~al.} 2004, \apjl,
  600, L107, \dodoi{10.1086/378578}

\bibitem[{{Ferreira} {et~al.}(2023){Ferreira}, {Conselice}, {Sazonova},
  {Ferrari}, {Caruana}, {Tohill}, {Lucatelli}, {Adams}, {Irodotou}, {Marshall},
  {Roper}, {Lovell}, {Verma}, {Austin}, {Trussler}, \& {Wilkins}}]{Ferreira23}
{Ferreira}, L., {Conselice}, C.~J., {Sazonova}, E., {et~al.} 2023, \apj, 955,
  94, \dodoi{10.3847/1538-4357/acec76}

\bibitem[{{Finkelstein} {et~al.}(2022){Finkelstein}, {Bagley}, {Haro},
  {Dickinson}, {Ferguson}, {Kartaltepe}, {Papovich}, {Burgarella}, {Kocevski},
  {Huertas-Company}, {Iyer}, {Koekemoer}, {Larson}, {P{\'e}rez-Gonz{\'a}lez},
  {Rose}, {Tacchella}, {Wilkins}, {Chworowsky}, {Medrano}, {Morales},
  {Somerville}, {Yung}, {Fontana}, {Giavalisco}, {Grazian}, {Grogin}, {Kewley},
  {Kirkpatrick}, {Kurczynski}, {Lotz}, {Pentericci}, {Pirzkal}, {Ravindranath},
  {Ryan}, {Trump}, {Yang}, {Almaini}, {Amor{\'\i}n}, {Annunziatella},
  {Backhaus}, {Barro}, {Behroozi}, {Bell}, {Bhatawdekar}, {Bisigello}, {Bromm},
  {Buat}, {Buitrago}, {Calabr{\`o}}, {Casey}, {Castellano}, {Ch{\'a}vez Ortiz},
  {Ciesla}, {Cleri}, {Cohen}, {Cole}, {Cooke}, {Cooper}, {Cooray}, {Costantin},
  {Cox}, {Croton}, {Daddi}, {Dav{\'e}}, {de La Vega}, {Dekel}, {Elbaz},
  {Estrada-Carpenter}, {Faber}, {Fern{\'a}ndez}, {Finkelstein}, {Freundlich},
  {Fujimoto}, {Garc{\'\i}a-Argum{\'a}nez}, {Gardner}, {Gawiser},
  {G{\'o}mez-Guijarro}, {Guo}, {Hamblin}, {Hamilton}, {Hathi}, {Holwerda},
  {Hirschmann}, {Hutchison}, {Jaskot}, {Jha}, {Jogee}, {Juneau}, {Jung},
  {Kassin}, {Le Bail}, {Leung}, {Lucas}, {Magnelli}, {Mantha}, {Matharu},
  {McGrath}, {McIntosh}, {Merlin}, {Mobasher}, {Newman}, {Nicholls}, {Pandya},
  {Rafelski}, {Ronayne}, {Santini}, {Seill{\'e}}, {Shah}, {Shen}, {Simons},
  {Snyder}, {Stanway}, {Straughn}, {Teplitz}, {Vanderhoof}, {Vega-Ferrero},
  {Wang}, {Weiner}, {Willmer}, {Wuyts}, {Zavala}, \& {CEERS
  Team}}]{Finkesltein22}
{Finkelstein}, S.~L., {Bagley}, M.~B., {Haro}, P.~A., {et~al.} 2022, \apjl,
  940, L55, \dodoi{10.3847/2041-8213/ac966e}

\bibitem[{{Finkelstein} {et~al.}(2023){Finkelstein}, {Bagley}, {Ferguson},
  {Wilkins}, {Kartaltepe}, {Papovich}, {Yung}, {Arrabal Haro}, {Behroozi},
  {Dickinson}, {Kocevski}, {Koekemoer}, {Larson}, {Le Bail}, {Morales},
  {P{\'e}rez-Gonz{\'a}lez}, {Burgarella}, {Dav{\'e}}, {Hirschmann},
  {Somerville}, {Wuyts}, {Bromm}, {Casey}, {Fontana}, {Fujimoto}, {Gardner},
  {Giavalisco}, {Grazian}, {Grogin}, {Hathi}, {Hutchison}, {Jha}, {Jogee},
  {Kewley}, {Kirkpatrick}, {Long}, {Lotz}, {Pentericci}, {Pierel}, {Pirzkal},
  {Ravindranath}, {Ryan}, {Trump}, {Yang}, {Bhatawdekar}, {Bisigello}, {Buat},
  {Calabr{\`o}}, {Castellano}, {Cleri}, {Cooper}, {Croton}, {Daddi}, {Dekel},
  {Elbaz}, {Franco}, {Gawiser}, {Holwerda}, {Huertas-Company}, {Jaskot},
  {Leung}, {Lucas}, {Mobasher}, {Pandya}, {Tacchella}, {Weiner}, \&
  {Zavala}}]{Finkelstein23}
{Finkelstein}, S.~L., {Bagley}, M.~B., {Ferguson}, H.~C., {et~al.} 2023, \apjl,
  946, L13, \dodoi{10.3847/2041-8213/acade4}

\bibitem[{{Foreman-Mackey} {et~al.}(2013){Foreman-Mackey}, {Hogg}, {Lang}, \&
  {Goodman}}]{ForemanMackey13}
{Foreman-Mackey}, D., {Hogg}, D.~W., {Lang}, D., \& {Goodman}, J. 2013, \pasp,
  125, 306, \dodoi{10.1086/670067}

\bibitem[{{Furlong} {et~al.}(2017){Furlong}, {Bower}, {Crain}, {Schaye},
  {Theuns}, {Trayford}, {Qu}, {Schaller}, {Berthet}, \& {Helly}}]{Furlong17}
{Furlong}, M., {Bower}, R.~G., {Crain}, R.~A., {et~al.} 2017, \mnras, 465, 722,
  \dodoi{10.1093/mnras/stw2740}

\bibitem[{{Gaia Collaboration} {et~al.}(2021){Gaia Collaboration}, {Brown},
  {Vallenari}, {Prusti}, {de Bruijne}, {Babusiaux}, {Biermann}, {Creevey},
  {Evans}, {Eyer}, {Hutton}, {Jansen}, {Jordi}, {Klioner}, {Lammers},
  {Lindegren}, {Luri}, {Mignard}, {Panem}, {Pourbaix}, {Randich}, {Sartoretti},
  {Soubiran}, {Walton}, {Arenou}, {Bailer-Jones}, {Bastian}, {Cropper},
  {Drimmel}, {Katz}, {Lattanzi}, {van Leeuwen}, {Bakker}, {Cacciari},
  {Casta{\~n}eda}, {De Angeli}, {Ducourant}, {Fabricius}, {Fouesneau},
  {Fr{\'e}mat}, {Guerra}, {Guerrier}, {Guiraud}, {Jean-Antoine Piccolo},
  {Masana}, {Messineo}, {Mowlavi}, {Nicolas}, {Nienartowicz}, {Pailler},
  {Panuzzo}, {Riclet}, {Roux}, {Seabroke}, {Sordo}, {Tanga}, {Th{\'e}venin},
  {Gracia-Abril}, {Portell}, {Teyssier}, {Altmann}, {Andrae}, {Bellas-Velidis},
  {Benson}, {Berthier}, {Blomme}, {Brugaletta}, {Burgess}, {Busso}, {Carry},
  {Cellino}, {Cheek}, {Clementini}, {Damerdji}, {Davidson}, {Delchambre},
  {Dell'Oro}, {Fern{\'a}ndez-Hern{\'a}ndez}, {Galluccio}, {Garc{\'\i}a-Lario},
  {Garcia-Reinaldos}, {Gonz{\'a}lez-N{\'u}{\~n}ez}, {Gosset}, {Haigron},
  {Halbwachs}, {Hambly}, {Harrison}, {Hatzidimitriou}, {Heiter},
  {Hern{\'a}ndez}, {Hestroffer}, {Hodgkin}, {Holl}, {Jan{\ss}en}, {Jevardat de
  Fombelle}, {Jordan}, {Krone-Martins}, {Lanzafame}, {L{\"o}ffler}, {Lorca},
  {Manteiga}, {Marchal}, {Marrese}, {Moitinho}, {Mora}, {Muinonen}, {Osborne},
  {Pancino}, {Pauwels}, {Petit}, {Recio-Blanco}, {Richards}, {Riello},
  {Rimoldini}, {Robin}, {Roegiers}, {Rybizki}, {Sarro}, {Siopis}, {Smith},
  {Sozzetti}, {Ulla}, {Utrilla}, {van Leeuwen}, {van Reeven}, {Abbas}, {Abreu
  Aramburu}, {Accart}, {Aerts}, {Aguado}, {Ajaj}, {Altavilla}, {{\'A}lvarez},
  {{\'A}lvarez Cid-Fuentes}, {Alves}, {Anderson}, {Anglada Varela}, {Antoja},
  {Audard}, {Baines}, {Baker}, {Balaguer-N{\'u}{\~n}ez}, {Balbinot}, {Balog},
  {Barache}, {Barbato}, {Barros}, {Barstow}, {Bartolom{\'e}}, {Bassilana},
  {Bauchet}, {Baudesson-Stella}, {Becciani}, {Bellazzini}, {Bernet}, {Bertone},
  {Bianchi}, {Blanco-Cuaresma}, {Boch}, {Bombrun}, {Bossini}, {Bouquillon},
  {Bragaglia}, {Bramante}, {Breedt}, {Bressan}, {Brouillet}, {Bucciarelli},
  {Burlacu}, {Busonero}, {Butkevich}, {Buzzi}, {Caffau}, {Cancelliere},
  {C{\'a}novas}, {Cantat-Gaudin}, {Carballo}, {Carlucci}, {Carnerero},
  {Carrasco}, {Casamiquela}, {Castellani}, {Castro-Ginard}, {Castro Sampol},
  {Chaoul}, {Charlot}, {Chemin}, {Chiavassa}, {Cioni}, {Comoretto}, {Cooper},
  {Cornez}, {Cowell}, {Crifo}, {Crosta}, {Crowley}, {Dafonte}, {Dapergolas},
  {David}, {David}, {de Laverny}, {De Luise}, {De March}, {De Ridder}, {de
  Souza}, {de Teodoro}, {de Torres}, {del Peloso}, {del Pozo}, {Delbo},
  {Delgado}, {Delgado}, {Delisle}, {Di Matteo}, {Diakite}, {Diener},
  {Distefano}, {Dolding}, {Eappachen}, {Edvardsson}, {Enke}, {Esquej}, {Fabre},
  {Fabrizio}, {Faigler}, {Fedorets}, {Fernique}, {Fienga}, {Figueras},
  {Fouron}, {Fragkoudi}, {Fraile}, {Franke}, {Gai}, {Garabato},
  {Garcia-Gutierrez}, {Garc{\'\i}a-Torres}, {Garofalo}, {Gavras}, {Gerlach},
  {Geyer}, {Giacobbe}, {Gilmore}, {Girona}, {Giuffrida}, {Gomel}, {Gomez},
  {Gonzalez-Santamaria}, {Gonz{\'a}lez-Vidal}, {Granvik},
  {Guti{\'e}rrez-S{\'a}nchez}, {Guy}, {Hauser}, {Haywood}, {Helmi}, {Hidalgo},
  {Hilger}, {H{\l}adczuk}, {Hobbs}, {Holland}, {Huckle}, {Jasniewicz},
  {Jonker}, {Juaristi Campillo}, {Julbe}, {Karbevska}, {Kervella}, {Khanna},
  {Kochoska}, {Kontizas}, {Kordopatis}, {Korn}, {Kostrzewa-Rutkowska},
  {Kruszy{\'n}ska}, {Lambert}, {Lanza}, {Lasne}, {Le Campion}, {Le Fustec},
  {Lebreton}, {Lebzelter}, {Leccia}, {Leclerc}, {Lecoeur-Taibi}, {Liao},
  {Licata}, {Lindstr{\o}m}, {Lister}, {Livanou}, {Lobel}, {Madrero Pardo},
  {Managau}, {Mann}, {Marchant}, {Marconi}, {Marcos Santos}, {Marinoni},
  {Marocco}, {Marshall}, {Martin Polo}, {Mart{\'\i}n-Fleitas}, {Masip},
  {Massari}, {Mastrobuono-Battisti}, {Mazeh}, {McMillan}, {Messina},
  {Michalik}, {Millar}, {Mints}, {Molina}, {Molinaro}, {Moln{\'a}r},
  {Montegriffo}, {Mor}, {Morbidelli}, {Morel}, {Morris}, {Mulone}, {Munoz},
  {Muraveva}, {Murphy}, {Musella}, {Noval}, {Ord{\'e}novic}, {Orr{\`u}},
  {Osinde}, {Pagani}, {Pagano}, {Palaversa}, {Palicio}, {Panahi}, {Pawlak},
  {Pe{\~n}alosa Esteller}, {Penttil{\"a}}, {Piersimoni}, {Pineau}, {Plachy},
  {Plum}, {Poggio}, {Poretti}, {Poujoulet}, {Pr{\v{s}}a}, {Pulone}, {Racero},
  {Ragaini}, {Rainer}, {Raiteri}, {Rambaux}, {Ramos}, {Ramos-Lerate}, {Re
  Fiorentin}, {Regibo}, {Reyl{\'e}}, {Ripepi}, {Riva}, {Rixon}, {Robichon},
  {Robin}, {Roelens}, {Rohrbasser}, {Romero-G{\'o}mez}, {Rowell}, {Royer},
  {Rybicki}, {Sadowski}, {Sagrist{\`a} Sell{\'e}s}, {Sahlmann}, {Salgado},
  {Salguero}, {Samaras}, {Sanchez Gimenez}, {Sanna}, {Santove{\~n}a},
  {Sarasso}, {Schultheis}, {Sciacca}, {Segol}, {Segovia}, {S{\'e}gransan},
  {Semeux}, {Shahaf}, {Siddiqui}, {Siebert}, {Siltala}, {Slezak}, {Smart},
  {Solano}, {Solitro}, {Souami}, {Souchay}, {Spagna}, {Spoto}, {Steele},
  {Steidelm{\"u}ller}, {Stephenson}, {S{\"u}veges}, {Szabados}, {Szegedi-Elek},
  {Taris}, {Tauran}, {Taylor}, {Teixeira}, {Thuillot}, {Tonello}, {Torra},
  {Torra}, {Turon}, {Unger}, {Vaillant}, {van Dillen}, {Vanel}, {Vecchiato},
  {Viala}, {Vicente}, {Voutsinas}, {Weiler}, {Wevers}, {Wyrzykowski}, {Yoldas},
  {Yvard}, {Zhao}, {Zorec}, {Zucker}, {Zurbach}, \& {Zwitter}}]{Gaia21}
{Gaia Collaboration}, {Brown}, A.~G.~A., {Vallenari}, A., {et~al.} 2021, \aap,
  649, A1, \dodoi{10.1051/0004-6361/202039657}

\bibitem[{{Genel} {et~al.}(2018){Genel}, {Nelson}, {Pillepich}, {Springel},
  {Pakmor}, {Weinberger}, {Hernquist}, {Naiman}, {Vogelsberger}, {Marinacci},
  \& {Torrey}}]{Genel18}
{Genel}, S., {Nelson}, D., {Pillepich}, A., {et~al.} 2018, \mnras, 474, 3976,
  \dodoi{10.1093/mnras/stx3078}

\bibitem[{{Girelli} {et~al.}(2020){Girelli}, {Pozzetti}, {Bolzonella},
  {Giocoli}, {Marulli}, \& {Baldi}}]{Girelli20}
{Girelli}, G., {Pozzetti}, L., {Bolzonella}, M., {et~al.} 2020, \aap, 634,
  A135, \dodoi{10.1051/0004-6361/201936329}

\bibitem[{{Grogin} {et~al.}(2011){Grogin}, {Kocevski}, {Faber}, {Ferguson},
  {Koekemoer}, {Riess}, {Acquaviva}, {Alexander}, {Almaini}, {Ashby}, {Barden},
  {Bell}, {Bournaud}, {Brown}, {Caputi}, {Casertano}, {Cassata}, {Castellano},
  {Challis}, {Chary}, {Cheung}, {Cirasuolo}, {Conselice}, {Roshan Cooray},
  {Croton}, {Daddi}, {Dahlen}, {Dav{\'e}}, {de Mello}, {Dekel}, {Dickinson},
  {Dolch}, {Donley}, {Dunlop}, {Dutton}, {Elbaz}, {Fazio}, {Filippenko},
  {Finkelstein}, {Fontana}, {Gardner}, {Garnavich}, {Gawiser}, {Giavalisco},
  {Grazian}, {Guo}, {Hathi}, {H{\"a}ussler}, {Hopkins}, {Huang}, {Huang},
  {Jha}, {Kartaltepe}, {Kirshner}, {Koo}, {Lai}, {Lee}, {Li}, {Lotz}, {Lucas},
  {Madau}, {McCarthy}, {McGrath}, {McIntosh}, {McLure}, {Mobasher},
  {Moustakas}, {Mozena}, {Nandra}, {Newman}, {Niemi}, {Noeske}, {Papovich},
  {Pentericci}, {Pope}, {Primack}, {Rajan}, {Ravindranath}, {Reddy}, {Renzini},
  {Rix}, {Robaina}, {Rodney}, {Rosario}, {Rosati}, {Salimbeni}, {Scarlata},
  {Siana}, {Simard}, {Smidt}, {Somerville}, {Spinrad}, {Straughn}, {Strolger},
  {Telford}, {Teplitz}, {Trump}, {van der Wel}, {Villforth}, {Wechsler},
  {Weiner}, {Wiklind}, {Wild}, {Wilson}, {Wuyts}, {Yan}, \& {Yun}}]{Grogin11}
{Grogin}, N.~A., {Kocevski}, D.~D., {Faber}, S.~M., {et~al.} 2011, \apjs, 197,
  35, \dodoi{10.1088/0067-0049/197/2/35}

\bibitem[{{Guo} {et~al.}(2023){Guo}, {Jogee}, {Finkelstein}, {Chen}, {Wise},
  {Bagley}, {Barro}, {Wuyts}, {Kocevski}, {Kartaltepe}, {McGrath}, {Ferguson},
  {Mobasher}, {Giavalisco}, {Lucas}, {Zavala}, {Lotz}, {Grogin},
  {Huertas-Company}, {Vega-Ferrero}, {Hathi}, {Arrabal Haro}, {Dickinson},
  {Koekemoer}, {Papovich}, {Pirzkal}, {Yung}, {Backhaus}, {Bell},
  {Calabr{\`o}}, {Cleri}, {Coogan}, {Cooper}, {Costantin}, {Croton}, {Davis},
  {Dekel}, {Franco}, {Gardner}, {Holwerda}, {Hutchison}, {Pandya},
  {P{\'e}rez-Gonz{\'a}lez}, {Ravindranath}, {Rose}, {Trump}, {de la Vega}, \&
  {Wang}}]{Guo23}
{Guo}, Y., {Jogee}, S., {Finkelstein}, S.~L., {et~al.} 2023, \apjl, 945, L10,
  \dodoi{10.3847/2041-8213/acacfb}

\bibitem[{Harris {et~al.}(2020)Harris, Millman, van~der Walt, Gommers,
  Virtanen, Cournapeau, Wieser, Taylor, Berg, Smith, Kern, Picus, Hoyer, van
  Kerkwijk, Brett, Haldane, del R{\'{i}}o, Wiebe, Peterson,
  G{\'{e}}rard-Marchant, Sheppard, Reddy, Weckesser, Abbasi, Gohlke, \&
  Oliphant}]{harris20}
Harris, C.~R., Millman, K.~J., van~der Walt, S.~J., {et~al.} 2020, Nature, 585,
  357, \dodoi{10.1038/s41586-020-2649-2}

\bibitem[{{H{\"a}u{\ss}ler} {et~al.}(2013){H{\"a}u{\ss}ler}, {Bamford}, {Vika},
  {Rojas}, {Barden}, {Kelvin}, {Alpaslan}, {Robotham}, {Driver}, {Baldry},
  {Brough}, {Hopkins}, {Liske}, {Nichol}, {Popescu}, \& {Tuffs}}]{Haussler13}
{H{\"a}u{\ss}ler}, B., {Bamford}, S.~P., {Vika}, M., {et~al.} 2013, \mnras,
  430, 330, \dodoi{10.1093/mnras/sts633}

\bibitem[{{H{\"a}u{\ss}ler} {et~al.}(2022){H{\"a}u{\ss}ler}, {Vika}, {Bamford},
  {Johnston}, {Brough}, {Casura}, {Holwerda}, {Kelvin}, \&
  {Popescu}}]{Haussler22}
{H{\"a}u{\ss}ler}, B., {Vika}, M., {Bamford}, S.~P., {et~al.} 2022, \aap, 664,
  A92, \dodoi{10.1051/0004-6361/202142935}

\bibitem[{{Holwerda} {et~al.}(2015){Holwerda}, {Bouwens}, {Oesch}, {Smit},
  {Illingworth}, \& {Labbe}}]{Holwerda15}
{Holwerda}, B.~W., {Bouwens}, R., {Oesch}, P., {et~al.} 2015, \apj, 808, 6,
  \dodoi{10.1088/0004-637X/808/1/6}

\bibitem[{{Holwerda} {et~al.}(2023){Holwerda}, {Hsu}, {Hathi}, {Bisigello}, {de
  la Vega}, {Arrabal Haro}, {Bagley}, {Dickinson}, {Finkelstein}, {Kartaltepe},
  {Koekemoer}, {Papovich}, {Pirzkal}, {Cook}, {Robertson}, {Casey}, {Aganze},
  {P{\'e}rez-Gonz{\'a}lez}, {Lucas}, {Jogee}, {Wilkins}, {Burgarella}, \&
  {Kirkpatrick}}]{Holwerda23}
{Holwerda}, B.~W., {Hsu}, C.-C., {Hathi}, N., {et~al.} 2023, arXiv e-prints,
  arXiv:2309.05835, \dodoi{10.48550/arXiv.2309.05835}

\bibitem[{{Hoyos} {et~al.}(2011){Hoyos}, {den Brok}, {Verdoes Kleijn},
  {Carter}, {Balcells}, {Guzm{\'a}n}, {Peletier}, {Ferguson}, {Goudfrooij},
  {Graham}, {Hammer}, {Karick}, {Lucey}, {Matkovi{\'c}}, {Merritt}, {Mouhcine},
  \& {Valentijn}}]{Hoyos11}
{Hoyos}, C., {den Brok}, M., {Verdoes Kleijn}, G., {et~al.} 2011, \mnras, 411,
  2439, \dodoi{10.1111/j.1365-2966.2010.17855.x}

\bibitem[{{Hoyos} {et~al.}(2012){Hoyos}, {Arag{\'o}n-Salamanca}, {Gray},
  {Maltby}, {Bell}, {Barazza}, {B{\"o}hm}, {H{\"a}u{\ss}ler}, {Jahnke},
  {Jogee}, {Lane}, {McIntosh}, \& {Wolf}}]{Hoyos12}
{Hoyos}, C., {Arag{\'o}n-Salamanca}, A., {Gray}, M.~E., {et~al.} 2012, \mnras,
  419, 2703, \dodoi{10.1111/j.1365-2966.2011.19918.x}

\bibitem[{{Huang} {et~al.}(2017){Huang}, {Fall}, {Ferguson}, {van der Wel},
  {Grogin}, {Koekemoer}, {Lee}, {P{\'e}rez-Gonz{\'a}lez}, \& {Wuyts}}]{Huang17}
{Huang}, K.-H., {Fall}, S.~M., {Ferguson}, H.~C., {et~al.} 2017, \apj, 838, 6,
  \dodoi{10.3847/1538-4357/aa62a6}

\bibitem[{{Huertas-Company} {et~al.}(2023){Huertas-Company}, {Iyer},
  {Angeloudi}, {Bagley}, {Finkelstein}, {Kartaltepe}, {Sarmiento},
  {Vega-Ferrero}, {Arrabal Haro}, {Behroozi}, {Buitrago}, {Cheng}, {Costantin},
  {Dekel}, {Dickinson}, {Elbaz}, {Grogin}, {Hathi}, {Holwerda}, {Koekemoer},
  {Lucas}, {Papovich}, {P{\'e}rez-Gonz{\'a}lez}, {Pirzkal}, {Seill{\'e}}, {de
  la Vega}, {Wuyts}, {Yang}, \& {Yung}}]{HuertasCompany23}
{Huertas-Company}, M., {Iyer}, K.~G., {Angeloudi}, E., {et~al.} 2023, arXiv
  e-prints, arXiv:2305.02478, \dodoi{10.48550/arXiv.2305.02478}

\bibitem[{Hunter(2007)}]{Hunter07}
Hunter, J.~D. 2007, Computing in Science \& Engineering, 9, 90,
  \dodoi{10.1109/MCSE.2007.55}

\bibitem[{{Kartaltepe} {et~al.}(2023){Kartaltepe}, {Rose}, {Vanderhoof},
  {McGrath}, {Costantin}, {Cox}, {Yung}, {Kocevski}, {Wuyts}, {Ferguson},
  {Bagley}, {Finkelstein}, {Amor{\'\i}n}, {Andrews}, {Aarabal Haro},
  {Backhaus}, {Behroozi}, {Bisigello}, {Calabr{\`o}}, {Casey}, {Coogan},
  {Cooper}, {Croton}, {de la Vega}, {Dickinson}, {Fontana}, {Franco},
  {Grazian}, {Grogin}, {Hathi}, {Holwerda}, {Huertas-Company}, {Iyer}, {Jogee},
  {Jung}, {Kewley}, {Kirkpatrick}, {Koekemoer}, {Liu}, {Lotz}, {Lucas},
  {Newman}, {Pacifici}, {Pandya}, {Papovich}, {Pentericci},
  {P{\'e}rez-Gonz{\'a}lez}, {Petersen}, {Pirzkal}, {Rafelski}, {Ravindranath},
  {Simons}, {Snyder}, {Somerville}, {Stanway}, {Straughn}, {Tacchella},
  {Trump}, {Vega-Ferrero}, {Wilkins}, {Yang}, \& {Zavala}}]{Kartaltepe23}
{Kartaltepe}, J.~S., {Rose}, C., {Vanderhoof}, B.~N., {et~al.} 2023, \apjl,
  946, L15, \dodoi{10.3847/2041-8213/acad01}

\bibitem[{{Kauffmann} {et~al.}(2003){Kauffmann}, {Heckman}, {White}, {Charlot},
  {Tremonti}, {Brinchmann}, {Bruzual}, {Peng}, {Seibert}, {Bernardi},
  {Blanton}, {Brinkmann}, {Castander}, {Cs{\'a}bai}, {Fukugita}, {Ivezic},
  {Munn}, {Nichol}, {Padmanabhan}, {Thakar}, {Weinberg}, \&
  {York}}]{Kauffmann03}
{Kauffmann}, G., {Heckman}, T.~M., {White}, S. D.~M., {et~al.} 2003, \mnras,
  341, 33, \dodoi{10.1046/j.1365-8711.2003.06291.x}

\bibitem[{{Koekemoer} {et~al.}(2011){Koekemoer}, {Faber}, {Ferguson}, {Grogin},
  {Kocevski}, {Koo}, {Lai}, {Lotz}, {Lucas}, {McGrath}, {Ogaz}, {Rajan},
  {Riess}, {Rodney}, {Strolger}, {Casertano}, {Castellano}, {Dahlen},
  {Dickinson}, {Dolch}, {Fontana}, {Giavalisco}, {Grazian}, {Guo}, {Hathi},
  {Huang}, {van der Wel}, {Yan}, {Acquaviva}, {Alexander}, {Almaini}, {Ashby},
  {Barden}, {Bell}, {Bournaud}, {Brown}, {Caputi}, {Cassata}, {Challis},
  {Chary}, {Cheung}, {Cirasuolo}, {Conselice}, {Roshan Cooray}, {Croton},
  {Daddi}, {Dav{\'e}}, {de Mello}, {de Ravel}, {Dekel}, {Donley}, {Dunlop},
  {Dutton}, {Elbaz}, {Fazio}, {Filippenko}, {Finkelstein}, {Frazer}, {Gardner},
  {Garnavich}, {Gawiser}, {Gruetzbauch}, {Hartley}, {H{\"a}ussler},
  {Herrington}, {Hopkins}, {Huang}, {Jha}, {Johnson}, {Kartaltepe},
  {Khostovan}, {Kirshner}, {Lani}, {Lee}, {Li}, {Madau}, {McCarthy},
  {McIntosh}, {McLure}, {McPartland}, {Mobasher}, {Moreira}, {Mortlock},
  {Moustakas}, {Mozena}, {Nandra}, {Newman}, {Nielsen}, {Niemi}, {Noeske},
  {Papovich}, {Pentericci}, {Pope}, {Primack}, {Ravindranath}, {Reddy},
  {Renzini}, {Rix}, {Robaina}, {Rosario}, {Rosati}, {Salimbeni}, {Scarlata},
  {Siana}, {Simard}, {Smidt}, {Snyder}, {Somerville}, {Spinrad}, {Straughn},
  {Telford}, {Teplitz}, {Trump}, {Vargas}, {Villforth}, {Wagner}, {Wandro},
  {Wechsler}, {Weiner}, {Wiklind}, {Wild}, {Wilson}, {Wuyts}, \&
  {Yun}}]{Koekemoer11}
{Koekemoer}, A.~M., {Faber}, S.~M., {Ferguson}, H.~C., {et~al.} 2011, \apjs,
  197, 36, \dodoi{10.1088/0067-0049/197/2/36}

\bibitem[{{Kormendy} \& {Bender}(1996)}]{Kormendy96}
{Kormendy}, J., \& {Bender}, R. 1996, \apjl, 464, L119, \dodoi{10.1086/310095}

\bibitem[{{Kravtsov}(2013)}]{Kravtsov13}
{Kravtsov}, A.~V. 2013, \apjl, 764, L31, \dodoi{10.1088/2041-8205/764/2/L31}

\bibitem[{{Kriek} {et~al.}(2009){Kriek}, {van Dokkum}, {Labb{\'e}}, {Franx},
  {Illingworth}, {Marchesini}, \& {Quadri}}]{Kriek09}
{Kriek}, M., {van Dokkum}, P.~G., {Labb{\'e}}, I., {et~al.} 2009, \apj, 700,
  221, \dodoi{10.1088/0004-637X/700/1/221}

\bibitem[{{K{\"u}mmel} {et~al.}(2022){K{\"u}mmel}, {{\'A}lvarez-Ayll{\'o}n},
  {Bertin}, {Dubath}, {Gavazzi}, {Hartley}, \& {Schefer}}]{Kummel22}
{K{\"u}mmel}, M., {{\'A}lvarez-Ayll{\'o}n}, A., {Bertin}, E., {et~al.} 2022,
  arXiv e-prints, arXiv:2212.02428, \dodoi{10.48550/arXiv.2212.02428}

\bibitem[{{Labb{\'e}} {et~al.}(2023){Labb{\'e}}, {van Dokkum}, {Nelson},
  {Bezanson}, {Suess}, {Leja}, {Brammer}, {Whitaker}, {Mathews}, {Stefanon}, \&
  {Wang}}]{Labbe23}
{Labb{\'e}}, I., {van Dokkum}, P., {Nelson}, E., {et~al.} 2023, \nat, 616, 266,
  \dodoi{10.1038/s41586-023-05786-2}

\bibitem[{{Lang} {et~al.}(2014){Lang}, {Wuyts}, {Somerville}, {F{\"o}rster
  Schreiber}, {Genzel}, {Bell}, {Brammer}, {Dekel}, {Faber}, {Ferguson},
  {Grogin}, {Kocevski}, {Koekemoer}, {Lutz}, {McGrath}, {Momcheva}, {Nelson},
  {Primack}, {Rosario}, {Skelton}, {Tacconi}, {van Dokkum}, \&
  {Whitaker}}]{Lang14}
{Lang}, P., {Wuyts}, S., {Somerville}, R.~S., {et~al.} 2014, \apj, 788, 11,
  \dodoi{10.1088/0004-637X/788/1/11}

\bibitem[{{Lange} {et~al.}(2015){Lange}, {Driver}, {Robotham}, {Kelvin},
  {Graham}, {Alpaslan}, {Andrews}, {Baldry}, {Bamford}, {Bland-Hawthorn},
  {Brough}, {Cluver}, {Conselice}, {Davies}, {Haeussler}, {Konstantopoulos},
  {Loveday}, {Moffett}, {Norberg}, {Phillipps}, {Taylor},
  {L{\'o}pez-S{\'a}nchez}, \& {Wilkins}}]{Lange15}
{Lange}, R., {Driver}, S.~P., {Robotham}, A. S.~G., {et~al.} 2015, \mnras, 447,
  2603, \dodoi{10.1093/mnras/stu2467}

\bibitem[{{Long} {et~al.}(2023){Long}, {Antwi-Danso}, {Lambrides}, {Lovell},
  {de la Vega}, {Valentino}, {Zavala}, {Casey}, {Wilkins}, {Yung}, {Arrabal
  Haro}, {Bagley}, {Bisigello}, {Chworowsky}, {Cooper}, {Cooper}, {Cooray},
  {Croton}, {Dickinson}, {Finkelstein}, {Franco}, {Gould}, {Hirschmann},
  {Hutchison}, {Kartaltepe}, {Kocevski}, {Koekemoer}, {Lucas}, {McKinney},
  {Papovich}, {Perez-Gonzalez}, {Pirzkal}, \& {Santini}}]{Long23}
{Long}, A.~S., {Antwi-Danso}, J., {Lambrides}, E.~L., {et~al.} 2023, arXiv
  e-prints, arXiv:2305.04662, \dodoi{10.48550/arXiv.2305.04662}

\bibitem[{{Lovell} {et~al.}(2021){Lovell}, {Vijayan}, {Thomas}, {Wilkins},
  {Barnes}, {Irodotou}, \& {Roper}}]{Lovell21}
{Lovell}, C.~C., {Vijayan}, A.~P., {Thomas}, P.~A., {et~al.} 2021, \mnras, 500,
  2127, \dodoi{10.1093/mnras/staa3360}

\bibitem[{{Macci{\`o}} {et~al.}(2008){Macci{\`o}}, {Dutton}, \& {van den
  Bosch}}]{Maccio08}
{Macci{\`o}}, A.~V., {Dutton}, A.~A., \& {van den Bosch}, F.~C. 2008, \mnras,
  391, 1940, \dodoi{10.1111/j.1365-2966.2008.14029.x}

\bibitem[{{Margalef-Bentabol} {et~al.}(2016){Margalef-Bentabol}, {Conselice},
  {Mortlock}, {Hartley}, {Duncan}, {Ferguson}, {Dekel}, \&
  {Primack}}]{Margalef-Bentabol16}
{Margalef-Bentabol}, B., {Conselice}, C.~J., {Mortlock}, A., {et~al.} 2016,
  \mnras, 461, 2728, \dodoi{10.1093/mnras/stw1451}

\bibitem[{{Marinacci} {et~al.}(2018){Marinacci}, {Vogelsberger}, {Pakmor},
  {Torrey}, {Springel}, {Hernquist}, {Nelson}, {Weinberger}, {Pillepich},
  {Naiman}, \& {Genel}}]{Marinacci18}
{Marinacci}, F., {Vogelsberger}, M., {Pakmor}, R., {et~al.} 2018, \mnras, 480,
  5113, \dodoi{10.1093/mnras/sty2206}

\bibitem[{{Marshall} {et~al.}(2022){Marshall}, {Wilkins}, {Di Matteo}, {Roper},
  {Vijayan}, {Ni}, {Feng}, \& {Croft}}]{Marshall22}
{Marshall}, M.~A., {Wilkins}, S., {Di Matteo}, T., {et~al.} 2022, \mnras, 511,
  5475, \dodoi{10.1093/mnras/stac380}

\bibitem[{Martorano {et~al.}(2023)Martorano, van~der Wel, Bell, Franx,
  Whitaker, Nersesian, Price, Baes, Suess, Nelson, Miller, Bezanson, \&
  Brammer}]{Martorano23}
Martorano, M., van~der Wel, A., Bell, E.~F., {et~al.} 2023, Rest-Frame
  Near-Infrared Radial Light Profiles up to z=3 from JWST/NIRCam: Wavelength
  Dependence of the S\'ersic Index.
\newblock \doarXiv{2308.11392}

\bibitem[{{Matthee} {et~al.}(2023){Matthee}, {Naidu}, {Brammer}, {Chisholm},
  {Eilers}, {Goulding}, {Greene}, {Kashino}, {Labbe}, {Lilly}, {Mackenzie},
  {Oesch}, {Weibel}, {Wuyts}, {Xiao}, {Bordoloi}, {Bouwens}, {van Dokkum},
  {Illingworth}, {Kramarenko}, {Maseda}, {Mason}, {Meyer}, {Nelson}, {Reddy},
  {Shivaei}, {Simcoe}, \& {Yue}}]{Matthee23}
{Matthee}, J., {Naidu}, R.~P., {Brammer}, G., {et~al.} 2023, arXiv e-prints,
  arXiv:2306.05448, \dodoi{10.48550/arXiv.2306.05448}

\bibitem[{{Mo} {et~al.}(1998){Mo}, {Mao}, \& {White}}]{MMW98}
{Mo}, H.~J., {Mao}, S., \& {White}, S. D.~M. 1998, \mnras, 295, 319,
  \dodoi{10.1046/j.1365-8711.1998.01227.x}

\bibitem[{{Mosleh} {et~al.}(2012){Mosleh}, {Williams}, {Franx}, {Gonzalez},
  {Bouwens}, {Oesch}, {Labbe}, {Illingworth}, \& {Trenti}}]{Mosleh12}
{Mosleh}, M., {Williams}, R.~J., {Franx}, M., {et~al.} 2012, \apjl, 756, L12,
  \dodoi{10.1088/2041-8205/756/1/L12}

\bibitem[{{Moster} {et~al.}(2013){Moster}, {Naab}, \& {White}}]{Moster13}
{Moster}, B.~P., {Naab}, T., \& {White}, S. D.~M. 2013, \mnras, 428, 3121,
  \dodoi{10.1093/mnras/sts261}

\bibitem[{{Mowla} {et~al.}(2019){Mowla}, {van Dokkum}, {Brammer}, {Momcheva},
  {van der Wel}, {Whitaker}, {Nelson}, {Bezanson}, {Muzzin}, {Franx},
  {MacKenty}, {Leja}, {Kriek}, \& {Marchesini}}]{Mowla19}
{Mowla}, L.~A., {van Dokkum}, P., {Brammer}, G.~B., {et~al.} 2019, \apj, 880,
  57, \dodoi{10.3847/1538-4357/ab290a}

\bibitem[{{Naab} {et~al.}(2009){Naab}, {Johansson}, \& {Ostriker}}]{Naab09}
{Naab}, T., {Johansson}, P.~H., \& {Ostriker}, J.~P. 2009, \apjl, 699, L178,
  \dodoi{10.1088/0004-637X/699/2/L178}

\bibitem[{{Naiman} {et~al.}(2018){Naiman}, {Pillepich}, {Springel},
  {Ramirez-Ruiz}, {Torrey}, {Vogelsberger}, {Pakmor}, {Nelson}, {Marinacci},
  {Hernquist}, {Weinberger}, \& {Genel}}]{Naiman18}
{Naiman}, J.~P., {Pillepich}, A., {Springel}, V., {et~al.} 2018, \mnras, 477,
  1206, \dodoi{10.1093/mnras/sty618}

\bibitem[{{Nandra} {et~al.}(2015){Nandra}, {Laird}, {Aird}, {Salvato},
  {Georgakakis}, {Barro}, {Perez-Gonzalez}, {Barmby}, {Chary}, {Coil},
  {Cooper}, {Davis}, {Dickinson}, {Faber}, {Fazio}, {Guhathakurta}, {Gwyn},
  {Hsu}, {Huang}, {Ivison}, {Koo}, {Newman}, {Rangel}, {Yamada}, \&
  {Willmer}}]{Nandra15}
{Nandra}, K., {Laird}, E.~S., {Aird}, J.~A., {et~al.} 2015, \apjs, 220, 10,
  \dodoi{10.1088/0067-0049/220/1/10}

\bibitem[{{Nedkova} {et~al.}(2021){Nedkova}, {H{\"a}u{\ss}ler}, {Marchesini},
  {Dimauro}, {Brammer}, {Eigenthaler}, {Feinstein}, {Ferguson},
  {Huertas-Company}, {Johnston}, {Kado-Fong}, {Kartaltepe}, {Labb{\'e}},
  {Lange-Vagle}, {Martis}, {McGrath}, {Muzzin}, {Oesch}, {Ordenes-Brice{\~n}o},
  {Puzia}, {Shipley}, {Simmons}, {Skelton}, {Stefanon}, {van der Wel}, \&
  {Whitaker}}]{Nedkova21}
{Nedkova}, K.~V., {H{\"a}u{\ss}ler}, B., {Marchesini}, D., {et~al.} 2021,
  \mnras, 506, 928, \dodoi{10.1093/mnras/stab1744}

\bibitem[{{Nelson} {et~al.}(2018){Nelson}, {Pillepich}, {Springel},
  {Weinberger}, {Hernquist}, {Pakmor}, {Genel}, {Torrey}, {Vogelsberger},
  {Kauffmann}, {Marinacci}, \& {Naiman}}]{Nelson18}
{Nelson}, D., {Pillepich}, A., {Springel}, V., {et~al.} 2018, \mnras, 475, 624,
  \dodoi{10.1093/mnras/stx3040}

\bibitem[{{Newville} {et~al.}(2014){Newville}, {Stensitzki}, {Allen}, \&
  {Ingargiola}}]{Newville14}
{Newville}, M., {Stensitzki}, T., {Allen}, D.~B., \& {Ingargiola}, A. 2014,
  {LMFIT: Non-Linear Least-Square Minimization and Curve-Fitting for Python},
  0.8.0, Zenodo,  Zenodo, \dodoi{10.5281/zenodo.11813}

\bibitem[{{Oesch} {et~al.}(2010){Oesch}, {Bouwens}, {Carollo}, {Illingworth},
  {Trenti}, {Stiavelli}, {Magee}, {Labb{\'e}}, \& {Franx}}]{Oesch10}
{Oesch}, P.~A., {Bouwens}, R.~J., {Carollo}, C.~M., {et~al.} 2010, \apjl, 709,
  L21, \dodoi{10.1088/2041-8205/709/1/L21}

\bibitem[{{Oke} \& {Gunn}(1983)}]{Oke83}
{Oke}, J.~B., \& {Gunn}, J.~E. 1983, \apj, 266, 713, \dodoi{10.1086/160817}

\bibitem[{{Ono} {et~al.}(2013){Ono}, {Ouchi}, {Curtis-Lake}, {Schenker},
  {Ellis}, {McLure}, {Dunlop}, {Robertson}, {Koekemoer}, {Bowler}, {Rogers},
  {Schneider}, {Charlot}, {Stark}, {Shimasaku}, {Furlanetto}, \&
  {Cirasuolo}}]{Ono13}
{Ono}, Y., {Ouchi}, M., {Curtis-Lake}, E., {et~al.} 2013, \apj, 777, 155,
  \dodoi{10.1088/0004-637X/777/2/155}

\bibitem[{{Ormerod} {et~al.}(2023){Ormerod}, {Conselice}, {Adams}, {Harvey},
  {Austin}, {Trussler}, {Ferreira}, {Caruana}, {Lucatelli}, {Li}, \&
  {Roper}}]{Ormerod23}
{Ormerod}, K., {Conselice}, C.~J., {Adams}, N.~J., {et~al.} 2023, arXiv
  e-prints, arXiv:2309.04377, \dodoi{10.48550/arXiv.2309.04377}

\bibitem[{{Pandya} {et~al.}(2023){Pandya}, {Zhang}, {Huertas-Company}, {Iyer},
  {McGrath}, {Barro}, {Finkelstein}, {Kuemmel}, {Hartley}, {Ferguson},
  {Kartaltepe}, {Primack}, {Dekel}, {Faber}, {Koo}, {Bryan}, {Somerville},
  {Amorin}, {Arrabal Haro}, {Bagley}, {Bell}, {Bertin}, {Costantin}, {Dave},
  {Dickinson}, {Feldmann}, {Fontana}, {Gavazzi}, {Giavalisco}, {Grazian},
  {Grogin}, {Guo}, {Hahn}, {Holwerda}, {Kewley}, {Kirkpatrick}, {Koekemoer},
  {Lotz}, {Lucas}, {Pentericci}, {Perez-Gonzalez}, {Pirzkal}, {Kocevski},
  {Papovich}, {Ravindranath}, {Rose}, {Schefer}, {Simons}, {Straughn},
  {Tacchella}, {Trump}, {de la Vega}, {Wilkins}, {Wuyts}, {Yang}, \&
  {Yung}}]{Pandya23}
{Pandya}, V., {Zhang}, H., {Huertas-Company}, M., {et~al.} 2023, arXiv
  e-prints, arXiv:2310.15232.
\newblock \doarXiv{2310.15232}

\bibitem[{{Peng} {et~al.}(2002){Peng}, {Ho}, {Impey}, \& {Rix}}]{Peng02}
{Peng}, C.~Y., {Ho}, L.~C., {Impey}, C.~D., \& {Rix}, H.-W. 2002, \aj, 124,
  266, \dodoi{10.1086/340952}

\bibitem[{{Peng} {et~al.}(2010){Peng}, {Ho}, {Impey}, \& {Rix}}]{Peng10}
---. 2010, \aj, 139, 2097, \dodoi{10.1088/0004-6256/139/6/2097}

\bibitem[{{Perrin} {et~al.}(2014){Perrin}, {Sivaramakrishnan}, {Lajoie},
  {Elliott}, {Pueyo}, {Ravindranath}, \& {Albert}}]{Perrin14}
{Perrin}, M.~D., {Sivaramakrishnan}, A., {Lajoie}, C.-P., {et~al.} 2014, in
  Society of Photo-Optical Instrumentation Engineers (SPIE) Conference Series,
  Vol. 9143, Space Telescopes and Instrumentation 2014: Optical, Infrared, and
  Millimeter Wave, ed. J.~{Oschmann}, Jacobus~M., M.~{Clampin}, G.~G. {Fazio},
  \& H.~A. {MacEwen}, 91433X, \dodoi{10.1117/12.2056689}

\bibitem[{{Pillepich} {et~al.}(2018){Pillepich}, {Nelson}, {Hernquist},
  {Springel}, {Pakmor}, {Torrey}, {Weinberger}, {Genel}, {Naiman}, {Marinacci},
  \& {Vogelsberger}}]{Pillepich18}
{Pillepich}, A., {Nelson}, D., {Hernquist}, L., {et~al.} 2018, \mnras, 475,
  648, \dodoi{10.1093/mnras/stx3112}

\bibitem[{{Robertson} {et~al.}(2006){Robertson}, {Bullock}, {Cox}, {Di Matteo},
  {Hernquist}, {Springel}, \& {Yoshida}}]{Robertson06}
{Robertson}, B., {Bullock}, J.~S., {Cox}, T.~J., {et~al.} 2006, \apj, 645, 986,
  \dodoi{10.1086/504412}

\bibitem[{{Roper} {et~al.}(2022){Roper}, {Lovell}, {Vijayan}, {Marshall},
  {Irodotou}, {Kuusisto}, {Thomas}, \& {Wilkins}}]{Roper22}
{Roper}, W.~J., {Lovell}, C.~C., {Vijayan}, A.~P., {et~al.} 2022, \mnras, 514,
  1921, \dodoi{10.1093/mnras/stac1368}

\bibitem[{{Schaye} {et~al.}(2015){Schaye}, {Crain}, {Bower}, {Furlong},
  {Schaller}, {Theuns}, {Dalla Vecchia}, {Frenk}, {McCarthy}, {Helly},
  {Jenkins}, {Rosas-Guevara}, {White}, {Baes}, {Booth}, {Camps}, {Navarro},
  {Qu}, {Rahmati}, {Sawala}, {Thomas}, \& {Trayford}}]{Schaye15}
{Schaye}, J., {Crain}, R.~A., {Bower}, R.~G., {et~al.} 2015, \mnras, 446, 521,
  \dodoi{10.1093/mnras/stu2058}

\bibitem[{{Sersic}(1968)}]{Sersic68}
{Sersic}, J.~L. 1968, {Atlas de Galaxias Australes}

\bibitem[{{Shen} {et~al.}(2003){Shen}, {Mo}, {White}, {Blanton}, {Kauffmann},
  {Voges}, {Brinkmann}, \& {Csabai}}]{Shen03}
{Shen}, S., {Mo}, H.~J., {White}, S. D.~M., {et~al.} 2003, \mnras, 343, 978,
  \dodoi{10.1046/j.1365-8711.2003.06740.x}

\bibitem[{{Shibuya} {et~al.}(2015){Shibuya}, {Ouchi}, \&
  {Harikane}}]{Shibuya15}
{Shibuya}, T., {Ouchi}, M., \& {Harikane}, Y. 2015, \apjs, 219, 15,
  \dodoi{10.1088/0067-0049/219/2/15}

\bibitem[{{Skelton} {et~al.}(2014){Skelton}, {Whitaker}, {Momcheva}, {Brammer},
  {van Dokkum}, {Labb{\'e}}, {Franx}, {van der Wel}, {Bezanson}, {Da Cunha},
  {Fumagalli}, {F{\"o}rster Schreiber}, {Kriek}, {Leja}, {Lundgren}, {Magee},
  {Marchesini}, {Maseda}, {Nelson}, {Oesch}, {Pacifici}, {Patel}, {Price},
  {Rix}, {Tal}, {Wake}, \& {Wuyts}}]{Skelton14}
{Skelton}, R.~E., {Whitaker}, K.~E., {Momcheva}, I.~G., {et~al.} 2014, \apjs,
  214, 24, \dodoi{10.1088/0067-0049/214/2/24}

\bibitem[{{Springel} {et~al.}(2018){Springel}, {Pakmor}, {Pillepich},
  {Weinberger}, {Nelson}, {Hernquist}, {Vogelsberger}, {Genel}, {Torrey},
  {Marinacci}, \& {Naiman}}]{Springel18}
{Springel}, V., {Pakmor}, R., {Pillepich}, A., {et~al.} 2018, \mnras, 475, 676,
  \dodoi{10.1093/mnras/stx3304}

\bibitem[{{Stefanon} {et~al.}(2017){Stefanon}, {Yan}, {Mobasher}, {Barro},
  {Donley}, {Fontana}, {Hemmati}, {Koekemoer}, {Lee}, {Lee}, {Nayyeri}, {Peth},
  {Pforr}, {Salvato}, {Wiklind}, {Wuyts}, {Ashby}, {Castellano}, {Conselice},
  {Cooper}, {Cooray}, {Dolch}, {Ferguson}, {Galametz}, {Giavalisco}, {Guo},
  {Willner}, {Dickinson}, {Faber}, {Fazio}, {Gardner}, {Gawiser}, {Grazian},
  {Grogin}, {Kocevski}, {Koo}, {Lee}, {Lucas}, {McGrath}, {Nandra}, {Newman},
  \& {van der Wel}}]{Stefanon17}
{Stefanon}, M., {Yan}, H., {Mobasher}, B., {et~al.} 2017, \apjs, 229, 32,
  \dodoi{10.3847/1538-4365/aa66cb}

\bibitem[{{Suess} {et~al.}(2022){Suess}, {Bezanson}, {Nelson}, {Setton},
  {Price}, {van Dokkum}, {Brammer}, {Labb{\'e}}, {Leja}, {Miller}, {Robertson},
  {Wel}, {Weaver}, \& {Whitaker}}]{Suess22}
{Suess}, K.~A., {Bezanson}, R., {Nelson}, E.~J., {et~al.} 2022, \apjl, 937,
  L33, \dodoi{10.3847/2041-8213/ac8e06}

\bibitem[{{Sun} {et~al.}(2023){Sun}, {Ho}, {Zhuang}, {Ma}, {Chen}, \&
  {Li}}]{Sun23}
{Sun}, W., {Ho}, L.~C., {Zhuang}, M.-Y., {et~al.} 2023, arXiv e-prints,
  arXiv:2308.09076, \dodoi{10.48550/arXiv.2308.09076}

\bibitem[{{Tomassetti} {et~al.}(2016){Tomassetti}, {Dekel}, {Mandelker},
  {Ceverino}, {Lapiner}, {Faber}, {Kneller}, {Primack}, \&
  {Sai}}]{Tomassetti16}
{Tomassetti}, M., {Dekel}, A., {Mandelker}, N., {et~al.} 2016, \mnras, 458,
  4477, \dodoi{10.1093/mnras/stw606}

\bibitem[{{Trujillo} {et~al.}(2006){Trujillo}, {F{\"o}rster Schreiber},
  {Rudnick}, {Barden}, {Franx}, {Rix}, {Caldwell}, {McIntosh}, {Toft},
  {H{\"a}ussler}, {Zirm}, {van Dokkum}, {Labb{\'e}}, {Moorwood},
  {R{\"o}ttgering}, {van der Wel}, {van der Werf}, \& {van
  Starkenburg}}]{Trujillo06}
{Trujillo}, I., {F{\"o}rster Schreiber}, N.~M., {Rudnick}, G., {et~al.} 2006,
  \apj, 650, 18, \dodoi{10.1086/506464}

\bibitem[{{van der Wel} {et~al.}(2012){van der Wel}, {Bell}, {H{\"a}ussler},
  {McGrath}, {Chang}, {Guo}, {McIntosh}, {Rix}, {Barden}, {Cheung}, {Faber},
  {Ferguson}, {Galametz}, {Grogin}, {Hartley}, {Kartaltepe}, {Kocevski},
  {Koekemoer}, {Lotz}, {Mozena}, {Peth}, \& {Peng}}]{vdw12}
{van der Wel}, A., {Bell}, E.~F., {H{\"a}ussler}, B., {et~al.} 2012, \apjs,
  203, 24, \dodoi{10.1088/0067-0049/203/2/24}

\bibitem[{{van der Wel} {et~al.}(2014{\natexlab{a}}){van der Wel}, {Franx},
  {van Dokkum}, {Skelton}, {Momcheva}, {Whitaker}, {Brammer}, {Bell}, {Rix},
  {Wuyts}, {Ferguson}, {Holden}, {Barro}, {Koekemoer}, {Chang}, {McGrath},
  {H{\"a}ussler}, {Dekel}, {Behroozi}, {Fumagalli}, {Leja}, {Lundgren},
  {Maseda}, {Nelson}, {Wake}, {Patel}, {Labb{\'e}}, {Faber}, {Grogin}, \&
  {Kocevski}}]{vdw14}
{van der Wel}, A., {Franx}, M., {van Dokkum}, P.~G., {et~al.}
  2014{\natexlab{a}}, \apj, 788, 28, \dodoi{10.1088/0004-637X/788/1/28}

\bibitem[{{van der Wel} {et~al.}(2014{\natexlab{b}}){van der Wel}, {Chang},
  {Bell}, {Holden}, {Ferguson}, {Giavalisco}, {Rix}, {Skelton}, {Whitaker},
  {Momcheva}, {Brammer}, {Kassin}, {Martig}, {Dekel}, {Ceverino}, {Koo},
  {Mozena}, {van Dokkum}, {Franx}, {Faber}, \& {Primack}}]{vdw14b}
{van der Wel}, A., {Chang}, Y.-Y., {Bell}, E.~F., {et~al.} 2014{\natexlab{b}},
  \apjl, 792, L6, \dodoi{10.1088/2041-8205/792/1/L6}

\bibitem[{{Vega-Ferrero} {et~al.}(2023){Vega-Ferrero}, {Huertas-Company},
  {Costantin}, {P{\'e}rez-Gonz{\'a}lez}, {Sarmiento}, {Kartaltepe},
  {Pillepich}, {Bagley}, {Finkelstein}, {McGrath}, {Knapen}, {Arrabal Haro},
  {Bell}, {Buitrago}, {Calabr{\`o}}, {Dekel}, {Dickinson}, {Dom{\'\i}nguez
  S{\'a}nchez}, {Elbaz}, {Ferguson}, {Giavalisco}, {Holwerda}, {Kocesvski},
  {Koekemoer}, {Pandya}, {Papovich}, {Pirzkal}, {Primack}, \&
  {Yung}}]{VegaFerrero23}
{Vega-Ferrero}, J., {Huertas-Company}, M., {Costantin}, L., {et~al.} 2023,
  arXiv e-prints, arXiv:2302.07277, \dodoi{10.48550/arXiv.2302.07277}

\bibitem[{{Vika} {et~al.}(2013){Vika}, {Bamford}, {H{\"a}u{\ss}ler}, {Rojas},
  {Borch}, \& {Nichol}}]{Vika13}
{Vika}, M., {Bamford}, S.~P., {H{\"a}u{\ss}ler}, B., {et~al.} 2013, \mnras,
  435, 623, \dodoi{10.1093/mnras/stt1320}

\bibitem[{Virtanen {et~al.}(2020)Virtanen, Gommers, Oliphant, Haberland, Reddy,
  Cournapeau, Burovski, Peterson, Weckesser, Bright, {van der Walt}, Brett,
  Wilson, Millman, Mayorov, Nelson, Jones, Kern, Larson, Carey, Polat, Feng,
  Moore, {VanderPlas}, Laxalde, Perktold, Cimrman, Henriksen, Quintero, Harris,
  Archibald, Ribeiro, Pedregosa, {van Mulbregt}, \& {SciPy 1.0
  Contributors}}]{Scipy20}
Virtanen, P., Gommers, R., Oliphant, T.~E., {et~al.} 2020, Nature Methods, 17,
  261, \dodoi{10.1038/s41592-019-0686-2}

\bibitem[{{Vulcani} {et~al.}(2014){Vulcani}, {Bamford}, {H{\"a}u{\ss}ler},
  {Vika}, {Rojas}, {Agius}, {Baldry}, {Bauer}, {Brown}, {Driver}, {Graham},
  {Kelvin}, {Liske}, {Loveday}, {Popescu}, {Robotham}, \& {Tuffs}}]{Vulcani14}
{Vulcani}, B., {Bamford}, S.~P., {H{\"a}u{\ss}ler}, B., {et~al.} 2014, \mnras,
  441, 1340, \dodoi{10.1093/mnras/stu632}

\bibitem[{{Whitaker} {et~al.}(2011){Whitaker}, {Labb{\'e}}, {van Dokkum},
  {Brammer}, {Kriek}, {Marchesini}, {Quadri}, {Franx}, {Muzzin}, {Williams},
  {Bezanson}, {Illingworth}, {Lee}, {Lundgren}, {Nelson}, {Rudnick}, {Tal}, \&
  {Wake}}]{Whitaker11}
{Whitaker}, K.~E., {Labb{\'e}}, I., {van Dokkum}, P.~G., {et~al.} 2011, \apj,
  735, 86, \dodoi{10.1088/0004-637X/735/2/86}

\bibitem[{{Williams} {et~al.}(2009){Williams}, {Quadri}, {Franx}, {van Dokkum},
  \& {Labb{\'e}}}]{Williams09}
{Williams}, R.~J., {Quadri}, R.~F., {Franx}, M., {van Dokkum}, P., \&
  {Labb{\'e}}, I. 2009, \apj, 691, 1879, \dodoi{10.1088/0004-637X/691/2/1879}

\bibitem[{{Wuyts} {et~al.}(2011){Wuyts}, {F{\"o}rster Schreiber}, {van der
  Wel}, {Magnelli}, {Guo}, {Genzel}, {Lutz}, {Aussel}, {Barro}, {Berta},
  {Cava}, {Graci{\'a}-Carpio}, {Hathi}, {Huang}, {Kocevski}, {Koekemoer},
  {Lee}, {Le Floc'h}, {McGrath}, {Nordon}, {Popesso}, {Pozzi}, {Riguccini},
  {Rodighiero}, {Saintonge}, \& {Tacconi}}]{Wuyts11}
{Wuyts}, S., {F{\"o}rster Schreiber}, N.~M., {van der Wel}, A., {et~al.} 2011,
  \apj, 742, 96, \dodoi{10.1088/0004-637X/742/2/96}

\bibitem[{{Zhang} {et~al.}(2019){Zhang}, {Primack}, {Faber}, {Koo}, {Dekel},
  {Chen}, {Ceverino}, {Chang}, {Fang}, {Guo}, {Lin}, \& {Wel}}]{Zhang19}
{Zhang}, H., {Primack}, J.~R., {Faber}, S.~M., {et~al.} 2019, \mnras, 484,
  5170, \dodoi{10.1093/mnras/stz339}

\end{thebibliography}
\bibliographystyle{aasjournal}

\appendix 

\section{Comparison to Single-Band Measurements Using GALFIT}
\label{sec: apdx}

\begin{figure*}[t!]
\includegraphics[width=\textwidth]{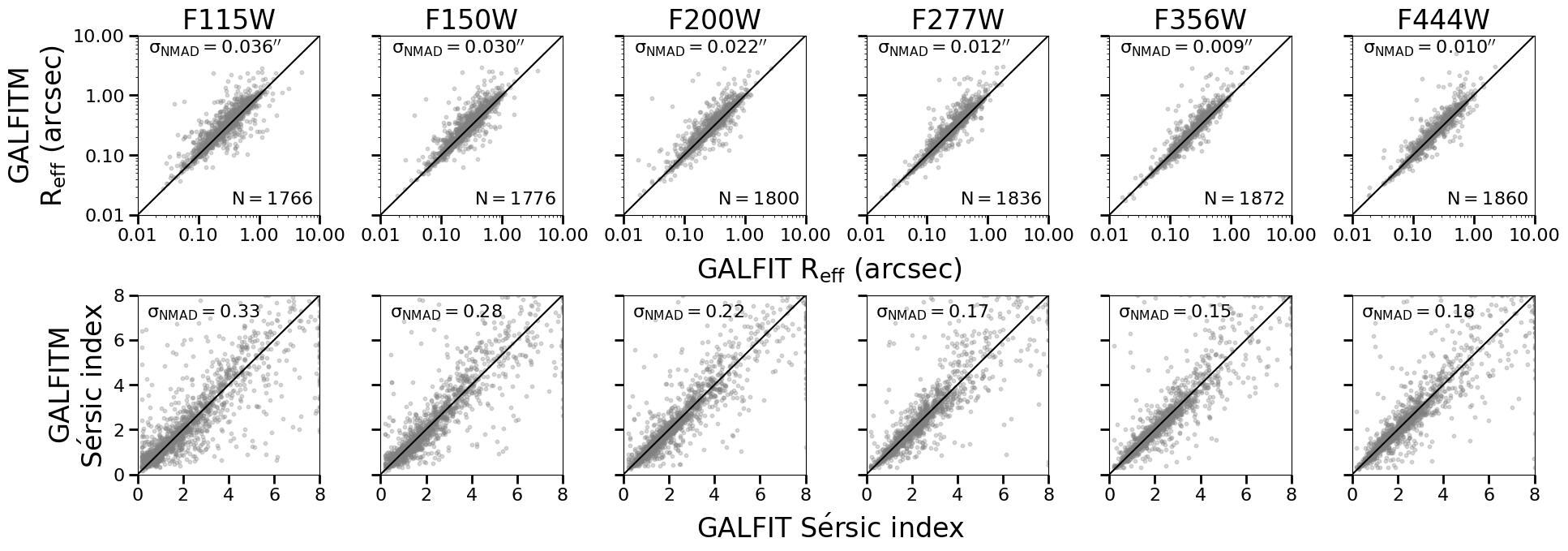}
\caption{Comparison of the sizes (top row) and the S\'{e}rsic indices (bottom row) between {\sc galfit} and {\sc galfitm}. We select $\sim1,800$ galaxies in our sample that have satisfactory {\sc galfit} fits (as discussed in the text). The number of galaxies that meet these fit criteria in each bandpass is listed in the lower right corner in each panel in the top row. Overall, there is good agreement between the two sets of measurements, as quantified by the NMAD scatter, which is displayed in the top left corner of each panel. The scatter is wavelength dependent: it is largest in F115W (left columns) and smallest at longer wavelengths (rightmost columns), though it is generally small: $\sigma_{\textrm{NMAD}}\sim0.02$ arcsec in size and $\sigma_{\textrm{NMAD}}\sim0.2$ in S\'{e}rsic index. }
\label{fig:galfit}
\end{figure*}

%\begin{itemize}
%    \item Scatter plots showing sizes and Sersic indices measured by Liz vs. our measurements 
%    \item we find that, in general, sizes measured using GALFITM and those estimated using GALFIT agree, but there is a noticeable fraction of the sample that has larger sizes measured using GALFITM than those estimated using GALFIT
%    \item \textcolor{red}{see Viraj's paper for comparable results regarding size differences between SE++ and GALFIT (cutting this for time, possibly can add during referee-ing process if needed)}
%    \item A quick examination of F444W sizes from GALFIT and those from GALFITM, colored by difference in axis ratios, shows that the galaxies with larger sizes in GALFITM have larger differences in axis ratio. This suggests that, for some galaxies, there is a noticeable impact when holding the axis ratio fixed in all bandpasses vs. letting it vary as a function of wavelength
%\end{itemize}

%In the preparation of this manuscript, we found our work to be complementary to another author (McGrath in prep.) who graciously offered to share their catalog for comparison to our measurements. 

In Figure \ref{fig:galfit}, we compare effective radii and S\'{e}rsic indices from {\sc galfit} measurements by McGrath et al. (in prep.) with our measurements using {\sc galfitm} in six NIRCam bandpasses in the top and bottom rows, respectively. We compare these measurements for $\sim1,800$ galaxies in our sample that are selected to have {\sc galfit} flag values equal to zero in each bandpass. A flag value equal to zero denotes fits that satisfy the following: the {\sc galfit}-estimated magnitudes are consistent with those estimatd by Source Extractor; all parameters lie within their respective constraints; {\sc galfit} found best-fitting solutions; and the objects are located far from the edge of the detector. In the {\sc galfit} fits, the position angle and axis ratio are left to vary freely in each bandpass independently, whereas in our {\sc galfitm} fits, these two parameters are fixed to the same values in all bandpasses.

There is good agreement in general between our {\sc galfitm} measurements and those performed using {\sc galfit}. Most galaxies, shown as gray dots in the figure, lie on or near the the one-to-one line, displayed as a black line. To quantify the scatter between the two sets of estimates, we use the NMAD, mentioned previously in Section \ref{sec:sed_fitting}, which is shown in the top left corner of each panel. Effective radii agree well, and the $\sigma_{\textrm{NMAD}}\sim0.02$ arcsec, though this depends on wavelength: the scatter is about four times larger in F115W than in F444W, though the scatter is at most 1.2 pixels. For the S\'{e}rsic indices, we also find good agreement: the typical $\sigma_{\textrm{NMAD}}\sim0.2$. This scatter also depends on wavelength, with the scatter in F115W about twice that in F444W.

\end{document}